\documentclass{article}
\usepackage{multicol}
\usepackage{arxiv}
\usepackage{placeins}
\usepackage[utf8]{inputenc} 
\usepackage[T1]{fontenc}    
\usepackage{xcolor}
\usepackage{hyperref}       
\usepackage{url}            
\usepackage{booktabs}       
\usepackage{amsfonts}       
\usepackage{amsmath}
\usepackage{amssymb}
\usepackage{nicefrac}       
\usepackage{microtype}      
\usepackage{float}          
\usepackage{newunicodechar}
\newunicodechar{₃}{$_3$}
\usepackage{lipsum}		
\usepackage{graphicx}
\usepackage{csquotes}
\usepackage[american]{babel}
\usepackage{multirow}

\usepackage[
  backend=biber,
  bibstyle=phys,
  citestyle=numeric-comp,
  sorting=none,
  biblabel=brackets,
  articletitle=true,
  chaptertitle=false,
  pageranges=false,
  doi=false,
  eprint=false,
  url=false
]{biblatex}

\usepackage{doi} 
\title{Loop-level surrogate modeling of dopant-distribution effects in Ba(Zr,Ti)O$_3$}

\author{
  Heiko Röthl\textsuperscript{1},
  Elke Kraker\textsuperscript{1},
  Julien Magnien\textsuperscript{1},
  Manfred Mücke\textsuperscript{1},
  Florian Mayer\textsuperscript{1}\textsuperscript{$\dagger$} \\
  \textsuperscript{1}Materials Center Leoben Forschung GmbH, Vordernberger Straße 12, 8700 Leoben, Austria \\
  \texttt{\textsuperscript{$\dagger$}Corresponding Author: florian.mayer@mcl.at}
}

\renewcommand{\headeright}{}
\renewcommand{\undertitle}{}
\renewcommand{\shorttitle}{}

\hypersetup{
pdftitle={Loop-level surrogate modeling for dopant-distribution engineering in BZT},
pdfsubject={cond-mat.mtrl-sci},
pdfauthor={Heiko Röthl, Florian Mayer},
pdfkeywords={Ferroelectrics, Conditional Autoencoder, Effective Hamiltonians},
}

\newcommand{\rev}[1]{{\color{black}#1}}

\begin{document}
\maketitle

\begin{abstract}
Barium titanate–based perovskites are important candidates for lead-free dielectric and electromechanical technologies. In Zr-substituted BaTiO$_3$ (BZT), functional behavior is usually discussed in terms of the average Zr concentration, while the influence of dopant spatial distribution beyond average concentration is less understood and difficult to explore systematically. Here we present an accelerated materials-design workflow that links controlled dopant distributions to full field-driven response curves. We generate a broad set of Zr distributions spanning a continuum of nanoscale arrangements, with layers, rods, dots, and lamellae serving as representative end-member motifs, and encode each configuration using a compact, parametrized descriptor model. Effective-Hamiltonian molecular dynamics is used to compute polarization–electric-field and strain–field hysteresis loops, and we train a conditional autoencoder surrogate to predict complete loops directly from the distribution parameters. This surrogate enables rapid screening and dense, property-selective design maps at scales that are not feasible with direct simulations alone, and it supports targeted follow-up simulations in regions of interest. Using the predicted loop database, we screen the distribution space for multiple functional targets, including energy-storage performance, electromechanical response, and switching behavior, and identify the corresponding dopant distribution motif families. The resulting design maps show that dopant distribution is an independent tuning parameter that can strongly affect hysteresis behavior and loop-derived figures of merit: layer-like motifs, vertical lamellae, and nanoplate-like inclusions emerge in different performance regimes. More generally, predicting full response curves enables screening of other loop-derived targets and multi-objective design in substituted ferroelectrics.
\end{abstract}

\keywords{Ferroelectrics \and Substituted Barium Titanate \and Conditional Autoencoder \and Effective Hamiltonians}

\begin{multicols}{2}
\section{Introduction}
Ferroelectric materials~\cite{Cohen1992,Haertling1999} show a broad range of functional responses, including dielectric~\cite{Wang2021}, electromechanical~\cite{Haertling1999}, and electrocaloric effects~\cite{Scott2011,Marathe2016}, and therefore remain important for many technologies. A key advantage of perovskite ferroelectrics is their chemical and structural tunability: relatively small changes in composition, local order, or defects can strongly modify phase stability and field-driven switching. Among lead-free perovskites, BaTiO$_3$ (BT) remains a benchmark system because its phase sequence, switching behavior, and chemical tunability are particularly well established~\cite{Shrout2007,Roedel2009,Shvartsman2012}.
BT exhibits the well-known sequence of structural phases~\cite{Gigli2022} (cubic paraelectric $\rightarrow$ tetragonal $\rightarrow$ orthorhombic $\rightarrow$ rhombohedral), and its rich phase behavior already illustrates that functional response is governed not only by average composition but also by how local environments and correlations develop across length scales.

Chemical substitution on the A- or B-site is widely used to tune perovskite ferroelectrics~\cite{Shvartsman2012,Vignas2022,Nishimatsu2016,Mayer2022a,Mentzer2019,Akbarzadeh2012}. Most studies parameterize substitution primarily through the average dopant fraction, which shifts phase boundaries, modifies susceptibilities, and can induce relaxor-like behavior~\cite{Vignas2022}. At fixed composition, however, the same dopant content can be realized through very different real-space arrangements — random, clustered, compositionally modulated, layered, or lamellar — and these arrangements need not yield the same field-driven response. Layering, compositional modulation, and atomically sharp substitution profiles are already established design motifs in ferroelectric heterostructures and superlattices. However, a systematic and quantitative understanding of how concentration and distribution jointly shape field-driven response, and how distribution effects compare at a given dopant fraction, is still limited.

In this work we focus on Zr-substituted BaTiO$_3$, Ba(Zr$_x$Ti$_{1-x}$)O$_3$ (BZT)~\cite{Mentzer2019,Mayer2022a,Bellaiche2017,Akbarzadeh2012}. Substitution of Ti by Zr on the B-site modifies phase stability and functional behavior: increasing Zr content changes the BT phase sequence~\cite{Petzelt2021,Mayer2022b}, suppresses ferroelectric order, and eventually drives a crossover toward relaxor-like response~\cite{Petzelt2021,Mentzer2019,Mayer2022b,Mayer2023}. More broadly, prior work on complex perovskites has shown that the emergence of ferroelectric versus relaxor-like behavior is closely tied to local structural arrangement~\cite{Grinberg2007Intro}, and simulation studies have emphasized that dielectric response is highly sensitive to local heterogeneity and its dynamics~\cite{Grinberg2009Intro}. Recent effective-Hamiltonian work on BZT has further shown that Zr substitution can stabilize nontrivial topological polarization textures, underscoring that chemically controlled local structure can generate qualitatively distinct polarization states beyond trends set by average composition alone~\cite{Mayer2026Intro}. Most studies of BZT nonetheless treat the average Zr concentration as the main control variable. Here we shift the emphasis and ask how much additional control resides in the nanoscale spatial arrangement of Zr at fixed composition.

Answering this question requires two ingredients. First, the distribution space must be described in a controlled way so that different motifs can be compared quantitatively. Second, the resulting structure space must be explored at a scale that is large enough to reveal reliable trends and to support design decisions. This is difficult with direct atomistic sampling alone: the number of distinct dopant arrangements grows rapidly with system size and motif complexity, and many figures of merit are determined by full polarization--electric-field (P--E) and strain--field (S--E) response curves rather than by a small set of scalar parameters. Moreover, meaningful comparisons often require separating concentration-driven changes from distribution-driven changes, for example by analyzing distributions at similar Zr concentration (\textit{zcon}). For this reason, a loop-level description is particularly useful: once full response curves are available, energy-storage measures, electromechanical indicators, and switching characteristics can all be extracted consistently, and multi-objective screening becomes possible when several targets must be balanced.

We therefore develop a simulation-driven surrogate workflow that makes dopant-distribution engineering computationally tractable. We first generate controlled Zr configurations in BT supercells, spanning layers, rods, dots, lamellae, and intermediate motifs, and describe them using a compact parametrized descriptor model inspired by geological pattern characterization. This provides a low-dimensional but physically interpretable representation of the distribution space, enabling systematic sampling and direct comparison across motif classes. We then compute the field-driven response of these configurations using first-principles-parameterized effective-Hamiltonian molecular dynamics (EH--MD), obtaining finite-temperature P--E and S--E hysteresis loops. To move beyond the directly simulated dataset, we train a conditional autoencoder surrogate~\cite{Butler2018,Schmidt2019,willard2022integrating} that predicts complete hysteresis responses from the distribution descriptors. The gain in screening efficiency is substantial: while generating on the order of $10^5$ loops with EH--MD would require roughly 2.7 million core-hours, the trained surrogate predicts the same number in minutes on a conventional computer. This makes it possible to construct dense design maps, screen for property-specific targets at scale, and select promising candidates for targeted EH--MD follow-up.

Machine learning has also been used to analyze ferroelectric switching and hysteretic responses in several related settings, including deep recurrent autoencoders for piezoresponse spectroscopy and encoder–decoder models linking domain structure to switching behavior~\cite{Kalinin2021,Agar2019}. More recently, Lahbabi et al.~\cite{Lahbabi2025} combined phase-field simulations and variational autoencoders to connect hysteresis-loop features to microstructural parameters in Hf$_{0.5}$Zr$_{0.5}$O$_2$. Relative to that literature, the present work targets a different design variable — dopant distribution in BZT — and conditions the surrogate directly on explicit real-space distribution descriptors to predict coupled loop-level responses.

Using BZT as a test case, we show that dopant distribution acts as an additional design variable that can strongly reshape hysteresis behavior and the resulting loop-derived figures of merit. Based on surrogate-generated design maps, we screen the descriptor space for energy-storage performance, electromechanical response, and switching characteristics, and identify the motif families associated with different response regimes. Although layered compositional modulation and related superlattice concepts are already well established in ferroelectrics~\cite{BENYOUSSEF2019279,CHAO2011978,Aramberri2022,Chen2023,Dimou2022}, the present study is not limited to that motif class. Its main contribution is a quantitative loop-level workflow that connects compact descriptors of dopant arrangement to full field-driven responses, thereby enabling systematic screening, interpretation, and targeted optimization across concentration and distribution variables.

\section{Methodology} \label{sec:methodology}
In this section, we introduce the simulation and modeling framework used to generate and analyze Zr-substituted BaTiO$_3$ supercells. We first describe the effective-Hamiltonian molecular-dynamics approach used to compute P--E and S--E responses. We then introduce the parametrized model used to generate controlled spatial distributions of Zr within the supercells, followed by a description of the dataset construction and the extraction of key functional quantities. Together, these components form the basis for the machine-learning framework developed in the subsequent section.

\subsection{Effective Hamiltonian Simulations} \label{sec:hamiltonian}
The atomistic response in this work is computed using molecular dynamics based on first-principles-derived effective Hamiltonians~\cite{Zhong1995,Rabe1995,Waghmare1997,Nishimatsu2008,Nishimatsu2010,Mayer2022b}, a framework that has been used extensively to describe finite-temperature behavior in ferroelectric perovskites and their chemically complex variants~\cite{Bellaiche2000Intro,Jorge2002Intro,Tinte2003Intro,Walizer2006Intro}. In this approach, the potential-energy surface is projected onto a reduced set of physically relevant lattice degrees of freedom~\cite{Zhong1995,Rabe1995}, most importantly local polar distortions and their couplings to strain, which allows large-scale finite-temperature simulations~\cite{Mayer2023} at tractable cost while retaining the key ingredients that govern ferroelectric switching and phase behavior~\cite{Paul2017,Mayer2022b}.

For pure BT we employ the effective-Hamiltonian parametrization of Mayer et al.~\cite{Mayer2022b}, which incorporates anharmonic couplings to additional optical modes~\cite{Mayer2022b,Paul2017} and reproduces the experimental transition sequence with improved accuracy~\cite{Mayer2022b}. For Ba(Zr$_x$Ti$_{1-x}$)O$_3$ we use the corresponding BZT extension introduced by the same authors~\cite{Mayer2022a}, which explicitly accounts for Zr substitution on the B site and reproduces the main features of the experimental composition–temperature phase behavior. All simulations were performed with a modified version of \textit{feram}~\cite{Nishimatsu2008,Nishimatsu2010,Nishimatsu2016}.

Because a central objective of this study is the evaluation of dynamic hysteresis behavior, we briefly summarize the key simulation parameters. All simulations were carried out at 300 K with a timestep of 2 fs. Prior to field cycling, systems were equilibrated for 200 ps. Polarization vs electric-field (P--E) and strain vs. electric-field (S--E) loops were obtained by applying an oscillating electric field along z-direction (pseudo-cubic [001] direction) with a maximum amplitude of 1500 kV/cm and a frequency of 1 GHz. Two full cycles were simulated: the first to eliminate transient effects and the second for data collection. The field-cycling protocol is chosen as a computationally consistent probe of relative hysteresis behavior across configurations rather than as a direct representation of experimental operating conditions. Temperature control was maintained using a velocity-scaling thermostat. All simulations employed a 40×40×40 (16 nm x 16 nm x 16 nm) supercell (described in detail in the following section) with fully periodic three-dimensional boundary conditions.

\subsection{Parametrized Zr Distribution Model} \label{sec:supercells}
As outlined in the introduction, a central objective of this work is to examine how different spatial arrangements of Zr ions within a BT matrix influence dielectric and electromechanical behavior. All simulations are based on $40\times40\times40$ supercells with periodic boundary conditions, providing 64,000 B-site lattice positions available for Ti$\rightarrow$Zr substitution. Even at fixed Zr concentration, the number of possible substitution patterns is combinatorially large, making an explicit enumeration infeasible. A controlled parametrization of spatial distributions is therefore essential to enable systematic exploration and subsequent machine-learning analysis.

To address this challenge, we introduce a constructive parametrization of an idealized but diverse motif space for Zr distributions, inspired by geological facies modeling~\cite{Deutsch2002,Chiles2012}. Accordingly, the present model focuses on controlled ordered motifs and their intermediates. Supercells are generated through a discrete layer-based procedure governed by a small set of tunable descriptors. Within this construction scheme, each three-dimensional Zr distribution is generated from five structural descriptors: \textit{zcon} (Zr concentration),  
\textit{intv} (number of active intervals),  
\textit{hff} (horizontal fill factor),  
\textit{hfr} (horizontal form ratio), and  
\textit{vsr} (vertical superposition ratio).

The overall Zr concentration (\textit{zcon}) is the fundamental parameter specifying the fraction of substituted B-sites and therefore constraining all other descriptors. Although it could range from 0 to 1, this study restricts \textit{zcon} to 0–0.3, where structural distribution effects remain pronounced; at higher concentrations the response increasingly approaches flattened P–E behavior with reduced configurational sensitivity.

The number of active vertical intervals (\textit{intv}) determines how Zr-containing layers are distributed along the supercell z-direction. An interval represents either a single active layer or a contiguous stack of active layers, and values between 1 and 40 are permitted. This parameter therefore controls the degree of vertical segmentation and the emergence of layered motifs.

Within each interval, the density of substitution is governed by the horizontal fill factor (\textit{hff}), which specifies the fraction of substituted sites per active layer. Larger \textit{hff} values yield denser layers and reduce the number of layers required to satisfy the global concentration constraint. Consequently, \textit{hff} mediates the trade-off between layer thickness and spatial extent of substitution.

The in-plane geometry of substituted regions is controlled by the horizontal form ratio (\textit{hfr}), which sets the aspect ratio of a continuous Zr-rich patch within each active layer. Values range from square-like domains to maximally elongated rectangular regions, enabling interpolation between isotropic and anisotropic substitution patterns.

Finally, the vertical superposition ratio (\textit{vsr}) introduces lateral translation between successive intervals. This parameter determines relative in-plane offsets between layers and thereby enables shifted stacking motifs under periodic boundary conditions, producing structures ranging from aligned nanolayers to staggered lamellae.
\rev{
To examine deviations from these sharply defined motifs, we additionally introduce an optional blurring parameter (\textit{rnd}). The parameter progressively redistributes Zr substitutions away from their original motif positions while preserving the total Zr concentration. Thus, \textit{rnd}=0 represents the unmodified parent structure, whereas increasing \textit{rnd} produces progressively more diffuse or disordered variants. The blurring parameter is not used as an input to the surrogate model, but is applied separately in selected EH--MD calculations discussed in Section~\ref{sec:random}.}

Representative distributions generated by this parametrization are shown in Figure~1. Panels (a–f) illustrate layered configurations obtained through controlled variation of individual parameters, including interval count, layer thickness, lateral extent, and inter-layer shifts. Additional examples demonstrate lamellar, rod-like, and dot-like motifs, highlighting that these recognizable structures represent limiting cases within a continuous design space rather than isolated constructions.
\rev{
Overall, the five primary descriptors (\textit{zcon}, \textit{intv}, \textit{hff}, \textit{hfr}, and \textit{vsr}) provide a compact and interpretable representation of the motif space suitable for systematic sampling and surrogate modeling. The optional blurring operation provides a density-preserving baseline for assessing the sensitivity of the identified motif--property relationships to substitutional disorder, but is not intended to reproduce a specific synthesis or interdiffusion mechanism. Additional algorithmic details are provided in the Supplementary Material~\cite{Supplemental}.}

\end{multicols}
\begin{figure}[H]
    \centering
    \includegraphics[width=0.90\linewidth]{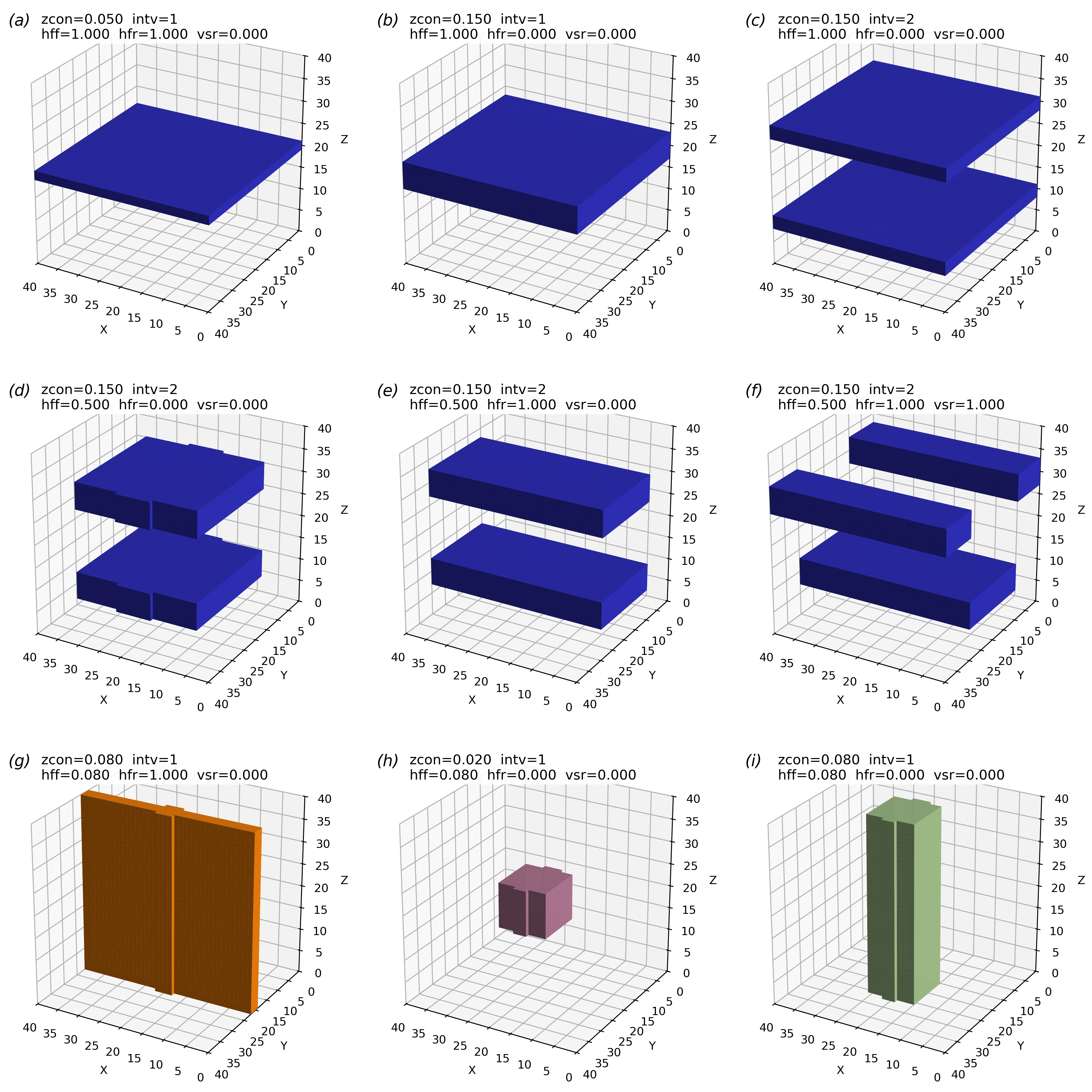}
    \caption{Examples of supercell configurations generated by the parameter-based modelling approach. Solid regions indicate cells where Ti is substituted by Zr. Panels (a)–(f) show a sequence of configurations obtained by varying only a single parameter between realizations. Panels (g)–(i) illustrate representative structures such as a nanolamella, a nanodot, and a nanorod, all produced using the same parametrized scheme.}
    \label{fig:supercelldispla}
\end{figure}
\begin{multicols}{2}

\subsection{Key Functional Quantities}
\begin{figure}[H]
    \centering
    \includegraphics[width=0.97\linewidth]{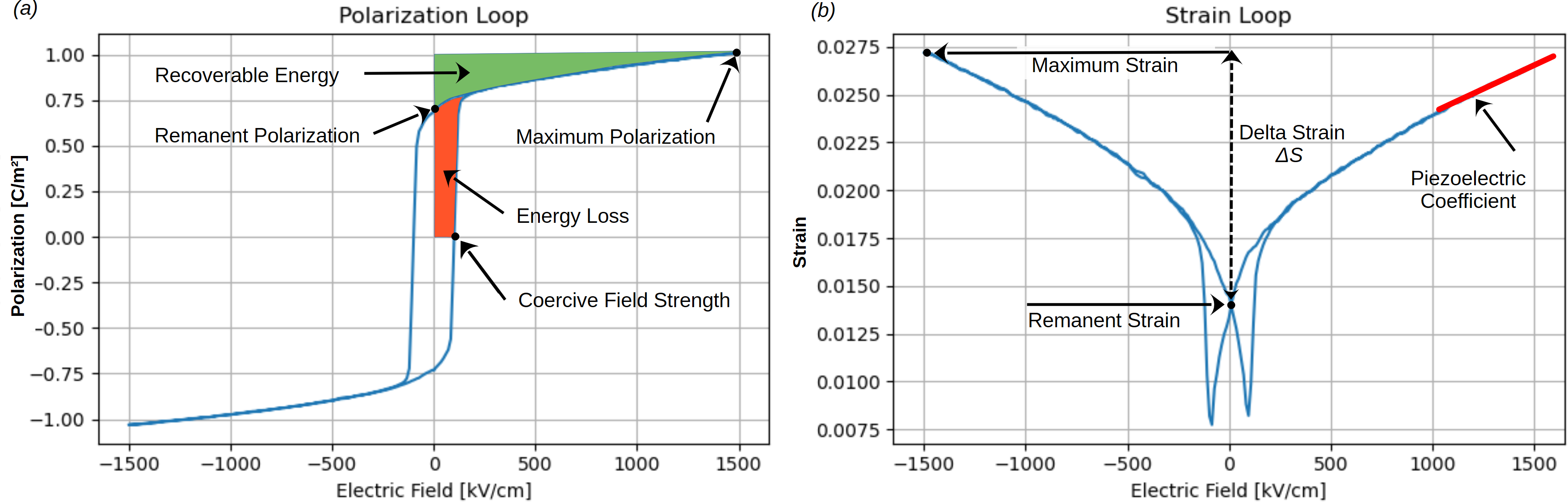}
    \caption{\rev{Schematic illustration of the key functional quantities extracted from the simulated hysteresis loops. (a) Polarization loop ($P$--$E$) defining the maximum polarization $P_{\max}$, remanent polarization $P_{\mathrm{rem}}$, and coercive field strength $E_c$. The shaded areas represent the recoverable energy density ($W_{\mathrm{rec}}$, green) and the energy loss ($W_{\mathrm{loss}}$, red). (b) Corresponding bipolar strain loop ($\eta_3$--$E$) detailing the maximum absolute strain, the remanent strain at zero field, and the resulting usable dynamic strain range ($\Delta \eta_3$ or Delta Strain $\Delta S$). The red solid line indicates the high-field linear fit used to evaluate the effective large-signal longitudinal piezoelectric coefficient ($d_{33}$).}}
    \label{fig:keyparams}
\end{figure}

We now briefly introduce the key functional quantities that will be discussed throughout the following sections. Several descriptors can be extracted directly from the P–E loops, including the maximum polarization $P_{\mathrm{max}}$, the remanent polarization $P_{\mathrm{rem}}$, and the coercive field $E_{\mathrm{c}}$. These quantities are illustrated in Figure~\ref{fig:keyparams} for clarity.

We are further interested in the energy-storage characteristics of the material, in particular the recoverable energy density 
$W_{rec}$ and the energy loss $W_{loss}$, which are standard figures of merit for dielectric capacitors~\cite{Li2021}. The recoverable energy density is computed according to Equation \ref{eq:wrec} and corresponds to the green-shaded area in Figure~\ref{fig:keyparams}. Here, $E_{dis}$ denotes the discharge field and $P$ the polarization. The losses are calculated using Equation \ref{eq:wloss}, which requires both the charging curve $E_{ch}$ and the discharging curve $E_{dis}$, the enclosed area between these curves, as illustrated in Figure~\ref{fig:keyparams}, represents $W_{loss}$. 

\begin{equation} \label{eq:wrec}
W_{\mathrm{rec}} 
= \int_{P_r}^{P_{\max}} E_{\mathrm{dis}}(P)\,\mathrm{d}P
\end{equation}

\begin{equation} \label{eq:wloss}
W_{\mathrm{loss}}
= \int_{0}^{P_{\max}} E_{\mathrm{ch}}(P)\,\mathrm{d}P 
 - \int_{P_r}^{P_{\max}} E_{\mathrm{dis}}(P)\,\mathrm{d}P
\end{equation}

From these equations, it becomes apparent that high energy losses are inherently associated with high remanent polarization and large coercive fields. Consequently, when the remanent polarization and coercive field are large, the recoverable energy density $W_{rec}$ is reduced, as a larger portion of the stored energy is dissipated. 

\rev{
In addition to dielectric properties, we evaluate the electromechanical performance via the bipolar strain loops ($\eta_3$--$E$ curves), as shown in Figure~\ref{fig:keyparams}(b). While the absolute maximum strain $\eta_{\max}$ describes the peak deformation at maximum field, practical actuator applications under cyclic loading are governed by the reversible electromechanical response. We therefore introduce the usable dynamic strain range, defined as $\Delta \eta_3 = \eta_{\max} - \eta_{\mathrm{rem}}$, where $\eta_{\mathrm{rem}}$ represents the remanent strain remaining at zero applied field after discharging. This subtraction effectively decouples the functional cyclic response from the static baseline strain shift induced by lattice expansion via Zr substitution.}

It is important to note that EH–MD simulations of this class are known to overestimate the electric-field scale by roughly one order of magnitude~\cite{Bellaiche2017}. The field-dependent quantities reported here should therefore be interpreted primarily in comparative rather than absolute terms. In the present study, the main emphasis is on relative trends, rankings, and motif-dependent differences across configurations evaluated under a common simulation protocol. Quantitatively accurate comparison to experiment would require explicit calibration of the field scale.

\end{multicols}

\begin{multicols}{2}

\section{Computational Experiments and Dataset}
Having defined the EH framework and the parametrized Zr-distribution model in Section 2, we now describe how these tools were combined to generate the dataset that underpins all subsequent analysis and machine-learning development. This section details how the structural parameter space was sampled, how EH--MD simulations were performed to obtain P--E and S--E responses, and how the resulting hysteresis loops were distilled into key functional quantities. Together, these computational experiments provide the quantitative foundation for training and validating the conditional autoencoder introduced in the following section.

\subsection{ Supercell sampling and MD simulations} \label{sec:supercells MD results}
Using the parametrized distribution model introduced in Section~\ref{sec:supercells}, we generated a dataset of 2,680 Zr-substituted BT supercells and evaluated their electromechanical response by EH–MD under the common protocol described in Section~\ref{sec:hamiltonian}. Dataset construction proceeded in three stages. First, 100 configurations were selected from the boundaries of the descriptor space to probe the range of accessible responses. Second, 1,000 additional configurations were sampled from the initial discrete grid of 7,000 parameter combinations. Third, the dataset was expanded to 2,680 configurations by generating off-grid samples that interpolate between the discrete descriptor values, thereby increasing coverage of the interior of the parameter space and reducing the bias associated with a purely grid-based design. \rev{Selected parent configurations were additionally evaluated at several blurring ratios using direct EH--MD simulations. These calculations were not used for surrogate training and are discussed separately in Section~5.3 to assess the influence of partial dopant disorder on the identified motif--property relationships.}

\begin{figure}[h]
    \centering
    \includegraphics[width=0.9\linewidth]{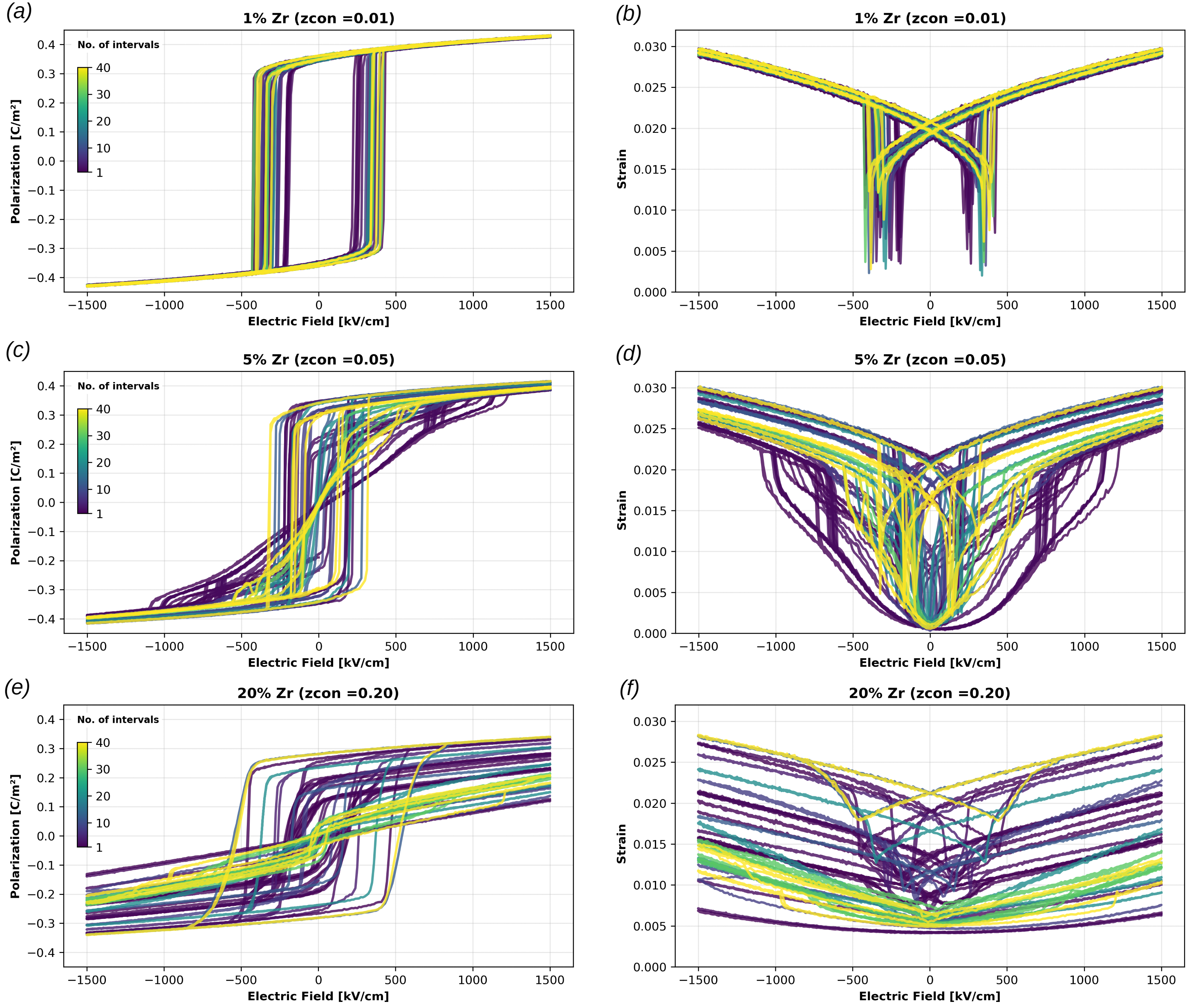}
    \caption{Polarization and strain–field loops for supercells with varying Zr distributions. Panels (a)–(b): 1\% Zr, showing minor variations mainly in coercive field; maximum polarization remains nearly constant. Panels (c)–(d): 5\% Zr, with more pronounced differences-some loops are ferroelectric, others flatter, approaching relaxor-like behavior. Panels (e)–(f): 20\% Zr, exhibiting substantial variability in loop shape, maximum polarization, coercive fields, and overall electromechanical response. The parametrized distribution approach captures a wide range of behavior across Zr concentrations.}
    \label{fig:initial_loops}
\end{figure}

Figure~\ref{fig:initial_loops} provides an overview of how different Zr distributions affect these hysteresis behaviors. Figures~\ref{fig:initial_loops}a and \ref{fig:initial_loops}b show loops for a low Zr concentration of 1\%. The color coding indicates the number of intervals (\textit{intv}). Even at 1\% Zr, noticeable variations appear, primarily in the coercive field. The maximum polarization remains nearly constant, which can be explained by the fact that Zr-substituted unit cells embedded in BT behave nearly nonpolar, preferring centrosymmetric positions~\cite{Mayer2022a,Mentzer2019}. Increasing Zr content therefore reduces the maximum polarization, although at 1\% this effect is still minimal. The strain–field loops also show small but distinct differences near the coercive field.

\end{multicols}
\begin{multicols}{2}

Figures \ref{fig:initial_loops}c and \ref{fig:initial_loops}d illustrate results for 5\% Zr. At this concentration, the impact of Zr arrangement becomes significantly more pronounced. Some configurations exhibit well-defined ferroelectric loops, while others show much flatter, weakly hysteretic responses reminiscent of relaxor-like behavior under the present simulation protocol. The maximum polarization remains relatively similar across configurations, as 5\% Zr still leads to only moderate suppression. The strain loops again display a wide range of responses.

Figures \ref{fig:initial_loops}e and \ref{fig:initial_loops}f present examples for 20\% Zr, where the variability becomes even more substantial. Differences appear not only in loop shape but also in maximum polarization, coercive fields, and the overall electromechanical response. The electromechanical response exhibits a markedly different behavior, which can be attributed to two primary factors. First, Zr substitution in BT induces a lattice expansion even in the absence of an electric field, leading to an increased baseline strain. Second, Zr can simultaneously suppress the ferroelectric and piezoelectric response. As a result, the strain under applied field may show only minimal variation or, conversely, a substantial change, depending on whether the Zr-induced change allows or hinders further field-driven response. 

Across the sampled configurations, \rev{the spread in coercive field, maximum polarization, and field-induced strain response increases strongly} with Zr concentration, confirming that distribution effects become more pronounced as the accessible motif space broadens.

Overall, these results demonstrate that even small concentrations of Zr can produce significant variations in dielectric and electromechanical behavior when the Zr ions are arranged differently. At higher concentrations, these effects become increasingly strong due to the greater structural degrees of freedom available. The final dataset therefore spans both descriptor-space extremes and interpolating interior configurations, providing a more suitable basis for surrogate training than a purely discrete grid.

\section{Machine Learning (ML) Model}
\label{sec:modelappraoch}

As outlined in the introduction, our goal is to develop a methodology that supports systematic exploration of dopant distributions in substituted systems for efficient materials design. While EH--MD is far more efficient than direct density-functional theory (DFT) and enables large-scale finite-temperature simulations, screening a large design space of dopant distributions with EH--MD alone still becomes computationally expensive. The parametrized generation of Zr distributions together with EH--MD thus provides an ideal foundation for applying machine-learning (ML) methods~\cite{Butler2018,Schmidt2019} to accelerate exploration and to identify promising regions of the design space. With suitable ML architectures, full response predictions can be obtained at a small fraction of the computational cost of MD, enabling high-throughput studies and rapid screening. All machine learning models  were implemented using the PyTorch deep learning framework~\cite{paszke2019pytorch}.

\subsection{General Aspects}
Each Zr distribution is defined by five structural descriptors: Zr concentration (\textit{zcon}), number of intervals (\textit{intv}), horizontal fill factor (\textit{hff}), horizontal form ratio (\textit{hfr}), and vertical superposition ratio (\textit{vsr}). Within the adopted construction scheme, these parameters determine the spatial arrangement of Zr within the 40×40×40 supercell. For every configuration, EH–MD simulations produce two response sequences: the polarization–electric field (P--E) loop and the strain–electric field (S--E) loop. From these sequences, functional descriptors such as recoverable energy density and piezoelectric response are derived. The dataset therefore consists of five structural inputs, two response sequences per configuration, and associated scalar descriptors.

From a materials-design perspective, the primary targets are scalar performance descriptors rather than the response sequences themselves. A direct regression from structural descriptors to these quantities is therefore a natural baseline, and in the present descriptor space it already captures much of the underlying structure--property relationship (see Supplementary Material~\cite{Supplemental}). However, direct scalar prediction does not provide access to the full hysteresis response and treats the derived quantities as independent outputs. We therefore adopt a loop-level surrogate strategy in which the model reconstructs the full P--E and S--E response sequences from the structural descriptors and derives the functional quantities only afterwards. This preserves the physical coupling embedded in the hysteresis behavior, enforces internal consistency between derived metrics, and allows additional descriptors to be evaluated without retraining. The resulting model is therefore not only predictive, but also more flexible as a framework for exploring the structural design space.

\subsection{Sequence Representation and Autoencoder Design}
To model electromechanical response, the P--E and S--E loops are treated as primary prediction targets. Autoencoders provide an effective framework for representing structured high-dimensional signals by encoding them into a compact latent representation and reconstructing them with minimal information loss~\cite{Hinton2006,Goodfellow2016}. In this context, the latent representation captures essential switching behavior, curvature, and coupling characteristics of the response sequences.

The encoder processes the sequences using one-dimensional convolutional layers followed by dense transformations. Convolutional operations capture local correlations along the field axis and efficiently represent switching features and loop curvature~\cite{LeCun2015}. The decoder mirrors this structure through dense expansion and transposed convolutions to recover the full sequences.

P--E and S--E loops are modeled jointly within a shared architecture. Since both responses originate from the same atomistic simulations, their coupled treatment encourages the learned representation to capture shared electromechanical structure rather than independent signal features. This formulation ensures that structural information directly influences sequence generation, enabling stable prediction for previously unsimulated configurations within the sampled descriptor space while maintaining a compact and interpretable representation. Additional details on architectural exploration, conditioning strategies, and training and optimization procedures are provided in the Supplementary Material~\cite{Supplemental}.

\subsection{Conditional Autoencoder Design}
To enable prediction of response sequences from structural descriptors, the autoencoder is extended to a conditional formulation in which reconstruction is guided by the distribution model parameters (\textit{zcon, intv, hff, hfr, vsr})~\cite{Mirza2014}. These descriptors are processed through a dedicated encoder branch and embedded into the latent representation during decoding, allowing the network to learn a unified structure–response mapping.

\end{multicols}
\begin{figure}[H]
    \centering
    \includegraphics[width=0.80\linewidth]{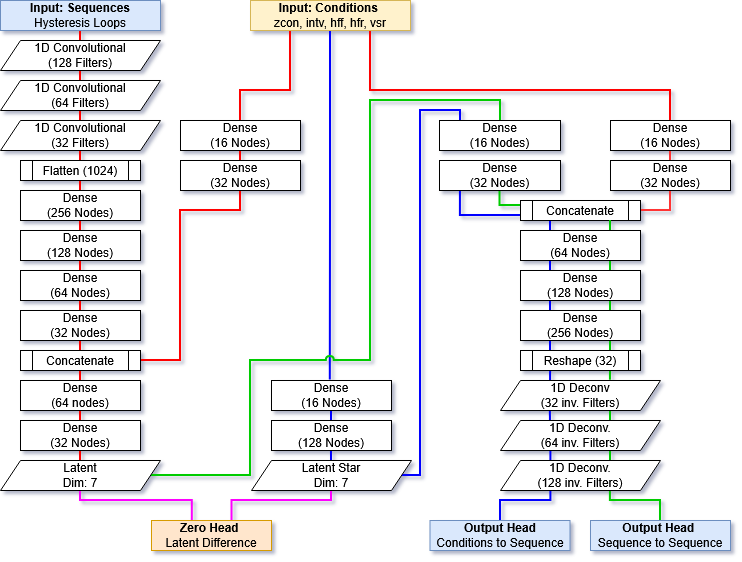}
    \caption{Schematic of the conditional autoencoder used to simultaneously predict  P–E and S–E  loops from structural parameters. The model has two inputs: the sequences themselves and the conditional data, corresponding to the supercell distribution parameters. The architecture features three output heads: two for reconstructing the sequences (sharing decoder weights) and one latent difference output head, a fixed-zero output that enforces alignment between the main latent space \textit{(Latent)} and the auxiliary latent space \textit{(Latent Star)}.}
    \label{fig:workflow}
\end{figure}
\vspace{-0.5em}
\begin{multicols}{2}

Unlike probabilistic generative formulations based on stochastic latent variables, such as variational autoencoders and their conditional extensions~\cite{NIPS2015_8d55a249}, the present deterministic conditional autoencoder targets reconstruction of configuration-specific physical responses, aligning directly with the objective of predicting unique hysteresis behavior.

The final architecture integrates sequence and conditional pathways within a single optimization framework. One encoder compresses the response sequences, while the second processes structural descriptors. Their latent representations are aligned through an auxiliary loss penalizing discrepancies between the two latent embeddings, encouraging consistency without overloading either pathway.

During decoding, conditional features are concatenated with latent sequence features and propagated through a shared decoder that produces both P–E and S–E outputs. Shared decoder weights enforce coherent reconstruction of the coupled responses and reduce the risk of divergence between predicted sequences. At inference time, only the conditional pathway is used: structural descriptors are mapped to the latent representation and decoded into predicted P–E and S–E loops.

This integrated formulation ensures that structural information directly influences sequence generation, enabling stable prediction for previously unsimulated configurations within the sampled descriptor space while maintaining a compact and interpretable latent representation. Further architectural details, ablation experiments, and alternative design variants explored during development are documented in the Supplementary Material~\cite{Supplemental}.

\subsection{Training and Performance Evaluation}
Training of the conditional autoencoder was carried out iteratively as the dataset expanded. An initial set of approximately 1,000 simulated P--E and S--E loops was used to establish the modeling workflow and refine the architecture. As additional configurations were generated through extended sampling of the structural parameter space, the dataset was progressively enlarged and the model retrained. The final network was trained on 2,680 loop pairs, ensuring that the learned representation captured the diversity of the explored configuration space.
Given the moderate dataset size relative to network capacity, model complexity was deliberately constrained through latent compression and weight sharing. These architectural choices act as implicit regularization mechanisms, limiting overfitting while preserving sufficient expressive power for accurate loop reconstruction.

Training employed a composite loss function that supervises all output branches simultaneously. The dominant contribution enforces accurate reconstruction of the P–E and S–E sequences, while auxiliary terms promote consistent encoding of the structural parameters and alignment of latent representations. Parameters were optimized using the Adam optimizer~\cite{Kingma2015}. Unless stated otherwise, model performance was evaluated using random training/validation/test partitions of the available dataset.

\begin{figure}[H]
    \centering
    \includegraphics[width=0.85\linewidth]{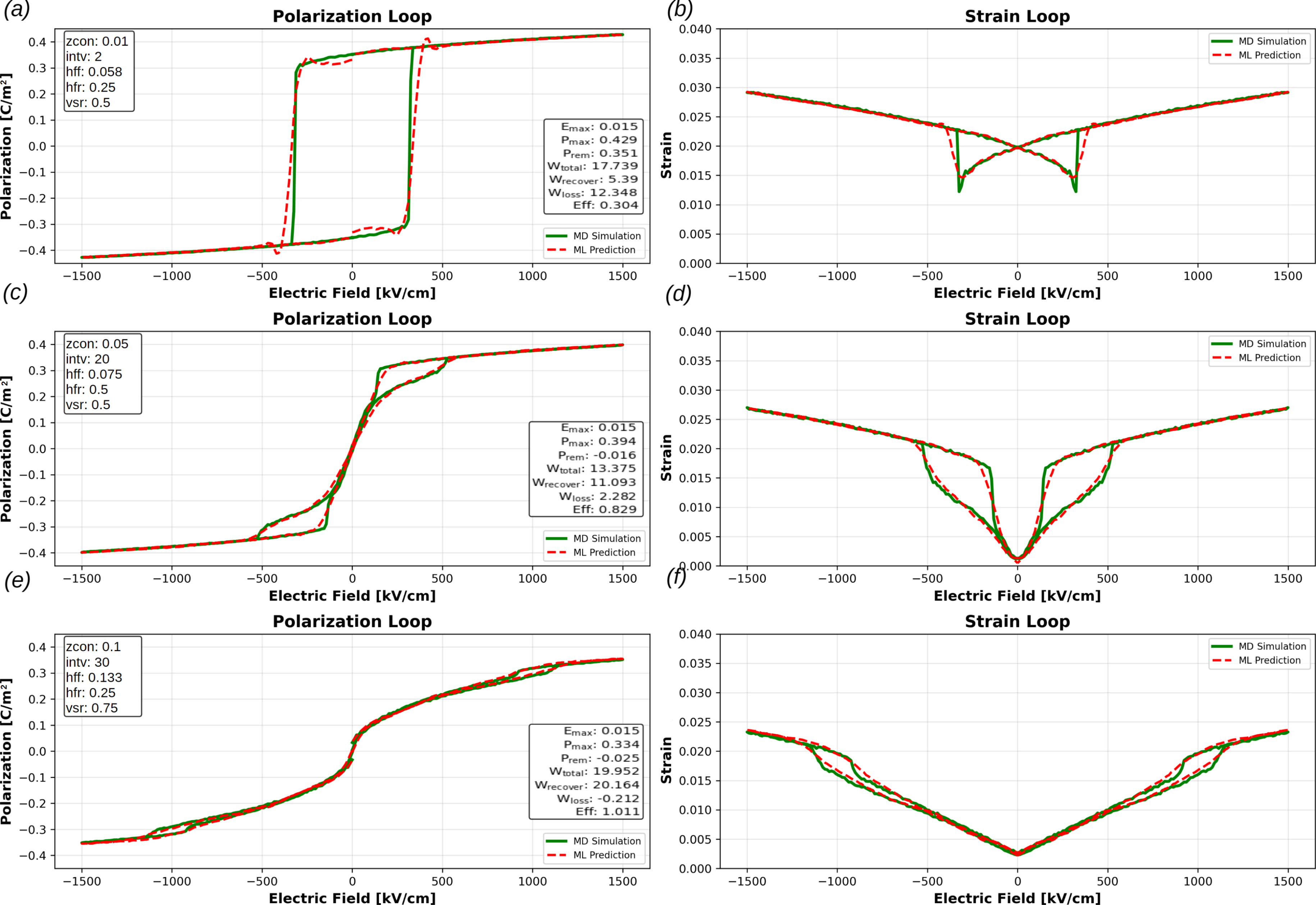}
    \caption{Comparison of EH--MD reference loops and cAE reconstructions for representative response regimes within the sampled descriptor space. Panels (a,b) show a strongly ferroelectric configuration with sharp switching and pronounced hysteresis in both the P--E and S--E responses. Panels (c,d) illustrate a flatter, weakly hysteretic double-loop-like response with a paraelectric-like central region and subsidiary branch structure. Panels (e,f) show a nearly linear, very weakly hysteretic response. Across these examples, the cAE reproduces the main loop morphology, including switching location, loop opening, curvature, and strain evolution.}
    \label{fig:samples}
\end{figure}

Overall convergence behavior and prediction accuracy indicate stable training dynamics and reliable reconstruction across the explored parameter space. Representative loop reconstructions (Figure~\ref{fig:samples}) illustrate the model’s ability to reproduce strongly ferroelectric, weakly hysteretic, and nearly linear responses across distinct regimes of loop morphology, supporting its use as a surrogate for large-scale screening of structural configurations. Further details on architectural evolution, loss-function variants, and extended performance evaluation are provided in the Supplementary Material~\cite{Supplemental}.

\end{multicols}
\begin{multicols}{2}

\subsection{Key Functional Quantities Prediction}
\begin{figure}[H]
    \centering
    \includegraphics[width=0.99\linewidth]{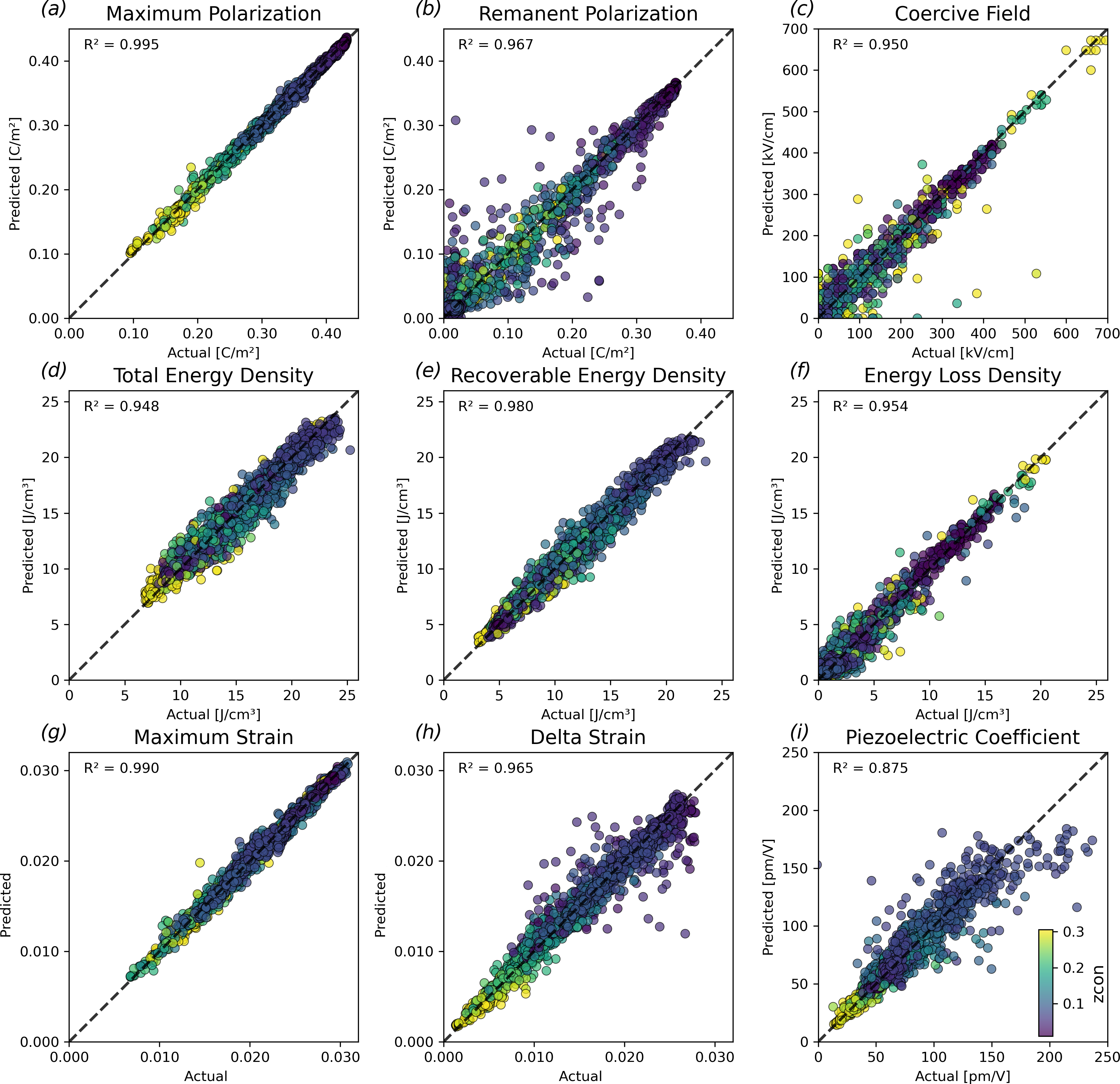}
\caption{Predicted versus reference values of loop-derived functional quantities for the conditional autoencoder surrogate. Panels (a)--(f) correspond to the electrical properties, including (a) maximum polarization $P_{\mathrm{max}}$, (b) remanent polarization $P_{\mathrm{rem}}$, (c) coercive field $E_{\mathrm{c}}$, (d) total energy density, (e) recoverable energy density $W_{\mathrm{rec}}$, and (f) energy loss density $W_{\mathrm{loss}}$. \rev{Panels (g)--(i) present the piezomechanical properties: (g) maximum strain $S_{\mathrm{max}}$, (h) recoverable strain $\Delta S$, and (i) the high-field effective slope denoted here as the piezoelectric coefficient $d_{33}$. Most quantities are reproduced with high fidelity, with $R^2 > 0.948$ for all metrics except $d_{33}$. The larger scatter in $P_{\mathrm{rem}}$ and especially $d_{33}$ reflects their stronger sensitivity to localized loop features, whereas integrated quantities such as $W_{\mathrm{rec}}$ and $W_{\mathrm{loss}}$ are less affected by small reconstruction errors. Colors indicate the Zr concentration ($zcon$).}}
\label{fig:pred-actula-xplot}
\end{figure}

Having established a stable and well-trained conditional autoencoder capable of accurately reproducing P--E and S--E loops, we now assess its ability to predict the derived functional quantities that ultimately define material performance. Because these descriptors — including recoverable energy density, remanent polarization, and piezoelectric response — are extracted from the reconstructed sequences rather than learned directly, their accuracy provides a stringent validation of the surrogate model. The predictive performance is summarized in Figure~\ref{fig:pred-actula-xplot}.

\end{multicols}
\vspace{-0.5em}
\begin{multicols}{2}

Overall, the agreement between predicted and reference values is strong. Coefficients of determination exceed 0.948 for all quantities except the piezoelectric coefficient, indicating that the model reliably captures both dielectric and electromechanical response characteristics across the explored parameter space.

Quantities derived from global loop features — namely total energy density (d), recoverable energy density (e), and energy loss (f) — are reproduced with particularly high fidelity. Because these metrics depend on integrated loop areas and discharge pathways, they are inherently smooth functionals of the sequences and therefore less sensitive to local reconstruction errors. Their tight correlation confirms the model’s accurate reproduction of overall loop geometry.

Descriptors determined by localized loop features exhibit slightly larger scatter. Remanent polarization (b), for example, is evaluated near steep and sometimes tilted loop segments; small deviations in reconstruction therefore translate into amplified variation in the extracted value. This effect remains moderate and does not alter the observed trends.
\rev{
The piezomechanical response quantities showcase the strength of the model across different strain-related features. The maximum strain $S_{max}$ (g) and the recoverable strain $\Delta S$ (h) both demonstrate excellent predictability. The largest dispersion appears for the piezoelectric coefficient $d_{33}$ (i).} This quantity is obtained from a numerical derivative evaluated only at the high-field tail of the S–E loop, making it highly sensitive to point-wise deviations and statistical noise. Despite this intrinsic sensitivity, the dominant trends remain well preserved.

Taken together, these results indicate that the surrogate is sufficiently accurate for ranking, screening, and motif-level interpretation across the sampled descriptor space, even though the most localized derivative-based quantities remain more sensitive to reconstruction error. To benchmark the loop-level surrogate against a simpler alternative, we additionally trained a standard multilayer perceptron (MLP) to predict the principal loop-derived scalar quantities directly from the five structural descriptors $(zcon, intv, hff, hfr, vsr)$. The MLP performs well across all targets, indicating that the adopted descriptor set already captures much of the relevant structure--property mapping within the sampled design space. However, the cAE provides the best overall performance for most quantities, with particularly clear improvements for remanent polarization and the energy-density-related metrics, while the MLP remains competitive and can match or slightly exceed the cAE for some localized targets on individual splits. A detailed comparison is provided in the Supplementary Material~\cite{Supplemental}. This baseline comparison shows that the principal advantage of the cAE is not only predictive accuracy, but also representational flexibility: unlike the direct scalar-regression baseline, it reconstructs complete P--E and S--E loops and therefore allows additional loop-derived figures of merit to be evaluated without retraining.

\section{Descriptor-based design maps for Zr distributions}
Now that a surrogate model is available that can predict complete P--E and S--E loops efficiently, we use it to explore the response landscape and to construct property-selective design maps that link Zr-distribution descriptors to loop-derived figures of merit. In the following, we consider three case studies that illustrate how the database can be screened for different functional targets. First, we analyze energy-storage performance and identify Zr distributions that combine high recoverable energy density with low hysteresis loss. Second, we examine electromechanical behavior using two contrasting objectives: \rev{configurations combining a high high-field effective slope $d_{33}$ with large recoverable strain $\Delta S$,} and mechanically quiet configurations with reduced electromechanical activity. Third, we analyze switching behavior by identifying distributions that yield easily switchable systems and, conversely, hard-switching systems. 
In the present manuscript, direct EH--MD follow-up for the energy-storage target is presented as the primary validation case in the main text, while additional EH--MD validation is included in the Supplementary Material~\cite{Supplemental}.

The results in this section should be interpreted mainly in terms of relative trends and rankings. As discussed above, effective-Hamiltonian simulations typically require applied electric fields that are about one order of magnitude larger than experimental values~\cite{Bellaiche2017}, and absolute energy densities are therefore expected to be higher than in experiment.

\subsection{Surrogate-generated loop database and screening protocol}

In this section we establish quantitative links between functional properties and the structural parameters of the parametrized Zr-distribution model. Each configuration is described by five distribution parameters (see Section~\ref{sec:supercells}). To identify correlations and to locate regions with favorable properties, we use the trained cAE to generate a dense prediction dataset. Specifically, we sample the descriptor space within the predefined parameter bounds (Section~\ref{sec:supercells}) and evaluate the cAE for 50{,}000 parameter sets. This yields a database of predicted P--E and S--E loops together with their associated distribution parameters. We then compute loop-derived key quantities from each predicted loop and add them to the database, which allows systematic screening by applying percentile-based filters to the derived metrics. The percentile thresholds used below are heuristic screening choices introduced to highlight the most informative regions for each case study.

\textbf{Energy-storage screening}

\begin{figure}[H]
\centering
\includegraphics[width=0.85\linewidth]{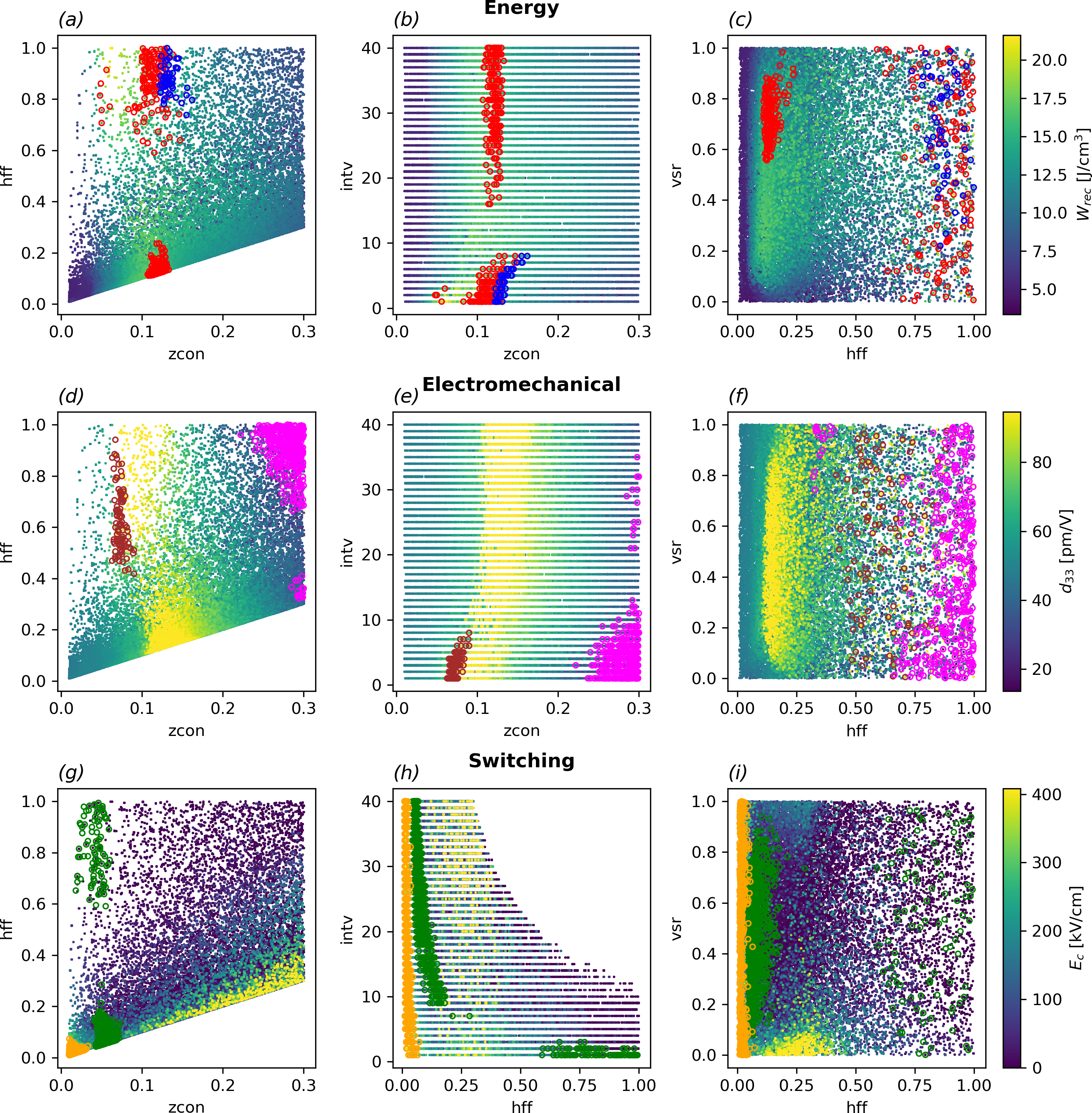}
\caption{Property-selective design maps derived from 50,000 cAE-predicted hysteresis loops. Each point corresponds to one predicted configuration and is plotted against selected distribution descriptors. In panels (a–c), red points highlight configurations with high recoverable energy density ($W_\mathrm{rec}\ge$ 90th percentile) and low hysteresis loss ($W_\mathrm{loss}\le$ 5th percentile), \rev{while blue points indicate a subset of these configurations that are additionally mechanically quiet ($\Delta S \le $ 30th percentile and $d_{33} \le $ 30th percentile). Panels (d–f) show electromechanical response screening using two complementary filters: high electromechanical activity (dark red points, $d_{33}\ge$ 85th percentile and $\Delta S \ge$ 85th percentile) and mechanically quiet response (magenta points, $d_{33}\le$ 2nd percentile and $\Delta S\le$ 2nd percentile). Panels (g–i) show switching-related screening: easy switching (green points, low $E_c$ and $P_r$ with high $P_\mathrm{max}$) and hard switching (orange points, high $E_c$ and $P_r$). A complete set of correlation plots for all descriptor pairs is provided in the Supplementary Material~\cite{Supplemental}.}}
\label{fig:correlation}
\end{figure}

We first use the database to identify configurations with favorable energy-storage properties. For this purpose, we apply a filter that selects configurations with high recoverable energy density ($W_\mathrm{rec} \ge$ 90th percentile) and low hysteresis loss ($W_\mathrm{loss} \le$ 5th percentile). The resulting subsets are highlighted in Figure~\ref{fig:correlation}. The first row of Figure~\ref{fig:correlation} relates $W_\mathrm{rec}$ to selected distribution parameters; a complete set of correlation plots is provided in the Supplementary Material~\cite{Supplemental}. In Figures~\ref{fig:correlation}a--c, the red points correspond to the energy-storage filter described above.

Because mechanical actuation can be undesirable in capacitor operation, we further identify a subset of mechanically quiet candidates. Starting from the red subset, we apply a second filter requiring low \rev{recoverable strain ($\Delta S \le$ 30th percentile)} and low high-field piezoelectric coefficient ($d_{33} \le$ 30th percentile). These doubly filtered configurations are highlighted in blue in Figures~\ref{fig:correlation}a--c. This second filter is relevant for capacitor applications where repeated field-induced strain could promote mechanical degradation.

\end{multicols}
\vspace{-0.5em}
\begin{multicols}{2}

Two main high-performance regions are visible for the energy-storage filter. Both occur in a similar Zr concentration range (\textit{zcon}), but they differ in other descriptors. In particular, Figure~\ref{fig:correlation}a indicates two red clusters with comparable \textit{zcon} but different horizontal fill factor (\textit{hff}). The same separation becomes clear in Figure~\ref{fig:correlation}b (similar \textit{zcon} but different interval counts) and Figure~\ref{fig:correlation}c (one cluster shows a relatively narrow range in \textit{vsr}, whereas the other spans a broader range).

\begin{figure}[H]
\centering
\includegraphics[width=0.85\linewidth]{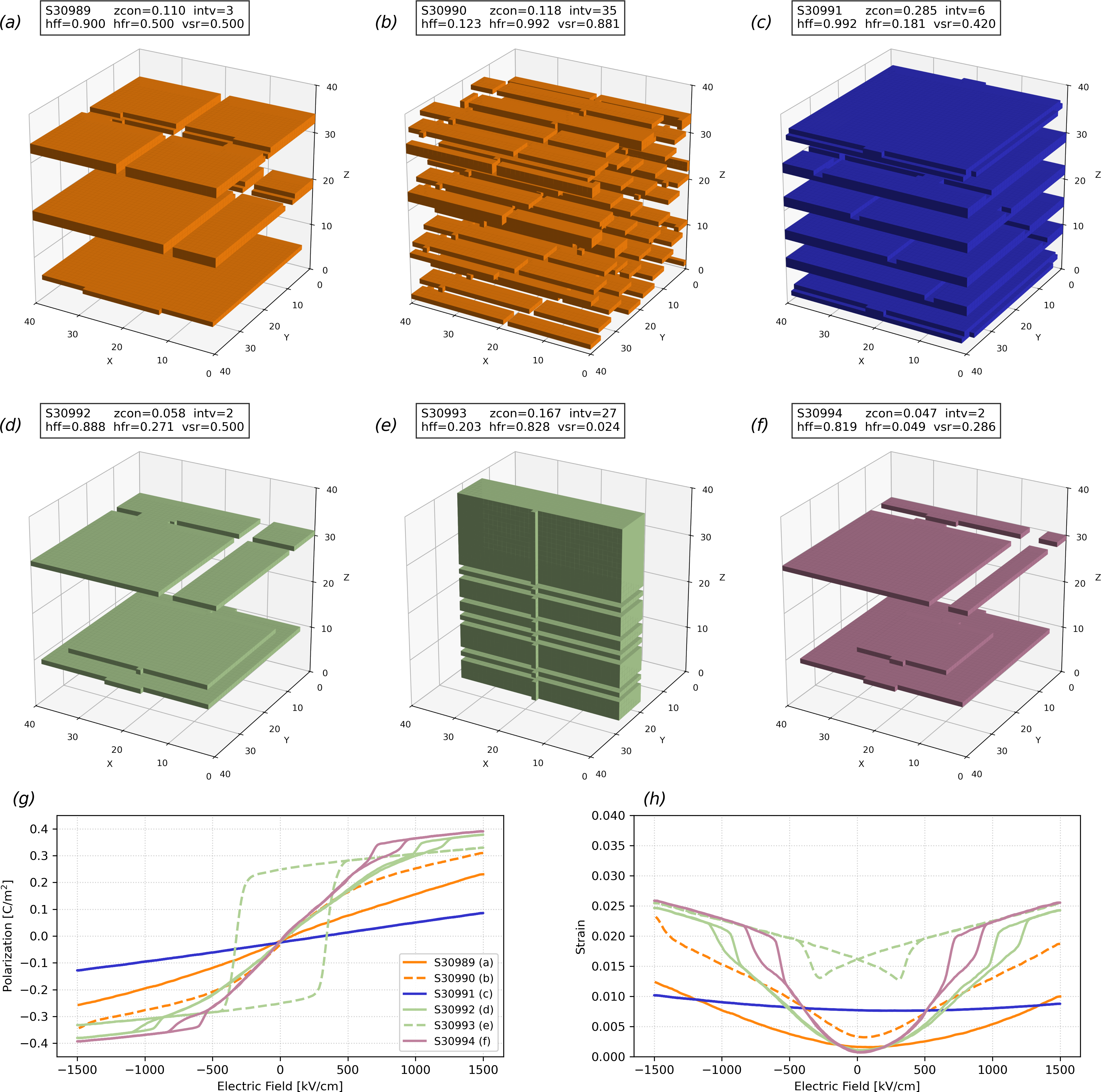}
\caption{Representative three-dimensional Zr-distribution motifs corresponding to the screened regions in Figure~\ref{fig:correlation}, along with their predicted functional responses. Panels (a) and (b) illustrate two high energy-storage motif families selected by the $W_\mathrm{rec}/W_\mathrm{loss}$ filter: (a) quasi-continuous Zr-rich nanolayers periodically interrupting BT-rich regions (superlattice-like modulation along $z$), and (b) distributed Zr-rich nanoplate-like inclusions embedded in the BT host. Panels (c)--(e) show representative electromechanical motifs: (c) a mechanically quiet layered configuration with small layer spacing, (d) a nanoplate-like configuration associated with large recoverable strain $\Delta S$ and a somewhat lower $d_{33}$, and (e) a vertical lamella-like motif combining moderate $\Delta S$ with a high high-field effective slope $d_{33}$.
Panel (f) shows a representative configuration from the easy-switching subset, illustrating that layer-like motifs can also produce slim loops with reduced coercive field and remanent polarization while maintaining high $P_{\text{max}}$. All structures are shown for the same supercell size. Zr-rich regions are highlighted for clarity. \rev{ Panels (g) and (h) display the corresponding cAE-predicted polarization ($P$--$E$) and strain ($S$--$E$) hysteresis loops, respectively, color-coded to match the individual structural configurations (a)–(f). The loop profiles directly illustrate how tailored spatial distribution motifs dictate macroscopic ferroelectric switching, energy-storage efficiency, and electromechanical strain curves.}}

\label{fig:3dstructures}
\end{figure}

These two regions correspond to two distinct motif families. The first family is characterized by large \textit{hff}, a small number of intervals, and a broader spread in \textit{vsr}. In real-space terms, these configurations correspond to quasi-continuous horizontal Zr-rich nanolayers that periodically interrupt BT-rich regions, i.e., a superlattice-like modulation along the field direction.An example is shown in Figure~\ref{fig:3dstructures}a,  \rev{with the corresponding P--E loop presented in Figure~\ref{fig:3dstructures}g.} Such layer-like architectures are associated with high recoverable energy and low loss, plausibly because the Zr-rich layers disrupt long-range ferroelectric correlations, which narrows the P--E loop and reduces hysteresis, while the BT-rich regions retain a high maximum polarization that supports a large energy density. This general mechanism is consistent with prior theoretical and experimental work on compositional modulation and superlattices~\cite{BENYOUSSEF2019279,CHAO2011978,Aramberri2022,Chen2023,Dimou2022}. The present analysis identifies the corresponding region in descriptor space and quantifies the associated trends. Notably, the blue subset largely overlaps with this first red cluster, indicating that these high-energy candidates can also be mechanically quiet and therefore attractive for capacitor operation.

The second energy-storage cluster shows lower \textit{hff}, a larger number of intervals, and a more concentrated \textit{vsr}. These configurations are different from extended nanolayers. A representative example is shown in Figure~\ref{fig:3dstructures}b, \rev{with the corresponding P--E loop presented in Figure~\ref{fig:3dstructures}g.} In real space, these motifs resemble distributed rectangular nanoplate-like Zr-rich regions embedded within BT. They are likewise consistent with a disruption of long-range ferroelectric correlations and can lead to narrower loops, while the surrounding BT-rich volume maintains a high $P_\mathrm{max}$. In this cluster, however, the mechanically quiet subset is less prominent, suggesting that the same energy-storage performance is achieved with somewhat stronger electromechanical activity.

\textbf{Electromechanical response screening}

We next examine electromechanical response using the same screening approach, not primarily to identify actuator materials, but to understand how field-induced strain activity varies across the descriptor space and how it overlaps with other functional targets, especially energy-storage behavior. To this end, we apply two complementary filters. The first selects configurations with strong electromechanical activity, defined here by high high-field effective slope $d_{33}$ and high \rev{recoverable strain} ($d_{33} \geq 85$th percentile and $\Delta S \geq 85$th percentile). The second selects mechanically quiet configurations with very low strain response ($d_{33} \leq 2$nd percentile and $\Delta S \leq 2$nd percentile). These subsets are shown in Figures~\ref{fig:correlation}d--f. The complete correlation structure and EH-MD validation are provided in the Supplementary Material~\cite{Supplemental}. \rev{As described in Section~\ref{sec:methodology}, $d_{33}$ is extracted from the high-field region of the S--E response and therefore represents a large-signal effective slope, whereas $\Delta S$ denotes the recoverable strain, defined as the difference between the maximum and remanent strain.}

For the high-electromechanical-activity subset, Figures~\ref{fig:correlation}d--f indicate two distinguishable regions. One occurs at somewhat higher \textit{zcon} and is concentrated at low \textit{hff}, while the other spans a broader range of \textit{hff}. For the low-\textit{hff} region (moderate \textit{zcon} combined with large \textit{hfr} and small \textit{vsr}), we observe vertically oriented lamella-like Zr-rich features that interrupt the BT host, as illustrated in Figure~\ref{fig:3dstructures}e. \rev{These configurations combine a high high-field effective slope $d_{33}$ with a moderate recoverable strain $\Delta S$. As shown by the corresponding S--E response in Figure~\ref{fig:3dstructures}h, the lamella-like structures exhibit an approximately linear strain evolution over a broad field range, which may be advantageous for proportional and predictable actuation. The second region is associated with nanoplate-like motifs (Figure~\ref{fig:3dstructures}d), which typically exhibit a larger $\Delta S$ but a somewhat smaller $d_{33}$. Their strain response is more strongly nonlinear, allowing a large recoverable strain to accumulate over the applied-field range. Thus, the two motif families offer complementary electromechanical characteristics: lamella-like structures favor a nearly linear response with a large effective slope, whereas nanoplate-like structures favor a larger recoverable strain.}

For the mechanically quiet subset, the filters identify configurations with simultaneously small recoverable $\Delta S$ and small high-field $d_{33}$. In the surrogate screening, these configurations fall predominantly into a layer-like motif family (Figure~\ref{fig:3dstructures}c), but with markedly smaller spacing between adjacent Zr-rich layers than in the high-electromechanical-activity layer case shown in Figure~\ref{fig:3dstructures}d. \rev{Correspondingly, the associated S--E response is nearly flat over the entire applied-field range, as shown in Figure~\ref{fig:3dstructures}h.} This highlights that ``layer-like'' is not sufficient as a descriptor on its own: layer spacing and thickness strongly affect whether layering promotes large strain response or suppresses it. This result is particularly relevant for the energy-storage screening, because it shows that mechanically quiet behavior can be obtained within the same motif class by tuning the geometric details of the layering, rather than requiring an entirely different type of distribution.

\textbf{Switching behavior screening}

Finally, we examine switching behavior by screening for configurations with easy versus hard switching. For easy switching, we select configurations with small coercive fields ($E_c \le$ 15th percentile), small remanent polarizations ($P_r \le$ 15th percentile), and simultaneously large maximum polarizations ($P_\mathrm{max} \ge$ 80th percentile). This combined filter targets loops that are slim and easy to switch while still maintaining a high polarization amplitude, which is of interest for energy storage. For hard switching, we select configurations with large coercive fields and remanent polarizations ($E_c \ge$ 95th percentile and $P_r \ge$ 95th percentile). The results are summarized in Figures~\ref{fig:correlation}g--i.

For the hard-switching filter, highlighted as orange points, the selected configurations lie, as expected, close to the BT-like regime. Pure BT exhibits a comparatively hard ferroelectric loop, and within the combined hard-switching criteria it appears as the most difficult-to-switch limit. If one considered only $E_c$ in isolation, other distributions could also yield elevated coercive fields, but the combined requirement of simultaneously large $E_c$ and large $P_r$ selects BT-like behavior most strongly.

\end{multicols}
\vspace{-0.5em}
\begin{multicols}{2}

For the easy-switching filter, marked as green points, the selected configurations overlap substantially with the layer-like motifs identified in the energy-storage screening. In particular, quasi-continuous Zr-rich nanolayers yield slim loops with reduced coercive field and remanent polarization while maintaining high $P_\mathrm{max}$, consistent with the requirement for easy switching and high energy density. A representative example is shown in Figure~\ref{fig:3dstructures}f, \rev{with the corresponding P--E and S--E loops presented in Figures~\ref{fig:3dstructures}g and \ref{fig:3dstructures}h, respectively.} Differences within this family are mainly quantitative, such as layer spacing, layer thickness, and the presence of small interruptions or secondary features, which can tune loop shape without changing the dominant motif class.

Overall, this section shows that cAE-based loop prediction combined with loop-derived key quantities enables systematic screening for target behavior and direct identification of the corresponding Zr-distribution motifs. Since the results presented here are based on surrogate predictions, we next validate the predicted trends and selected candidates by follow-up EH--MD simulations for representative configurations.

\subsection{Surrogate-guided validation and targeted sampling}
To translate the surrogate-derived design rules into validated microstructures, we combined rapid screening with targeted EH--MD simulations. Starting from the design maps, we carried out a denser screening in the region that combines high recoverable energy density with low hysteresis loss. The purpose of this step is to focus on configurations where the P--E loop becomes narrow while the polarization amplitude remains high, which is required for strong capacitor performance.

The main structural signatures of this region were identified from the previous analysis of the top-$W_\mathrm{rec}$ subset (Figure~\ref{fig:correlation}). High recoverable energy is associated with a narrow concentration window, quasi-continuous layering (large horizontal fill factor), and only a small number of Zr-rich intervals. Based on these signatures, we selected 300 candidate configurations and validated them by direct EH--MD simulations. For this selection, we constrained the three most influential parameters (\textit{zcon}, \textit{intv}, and \textit{hff}), while the remaining two parameters were sampled randomly. As a result, a small number of candidates fall outside the optimal region and show higher losses. These outliers are rare and mainly reflect unfavorable combinations of the unconstrained parameters; constraining all five parameters would further reduce this effect.

\begin{figure}[H]
\centering
\includegraphics[width=0.99\linewidth]{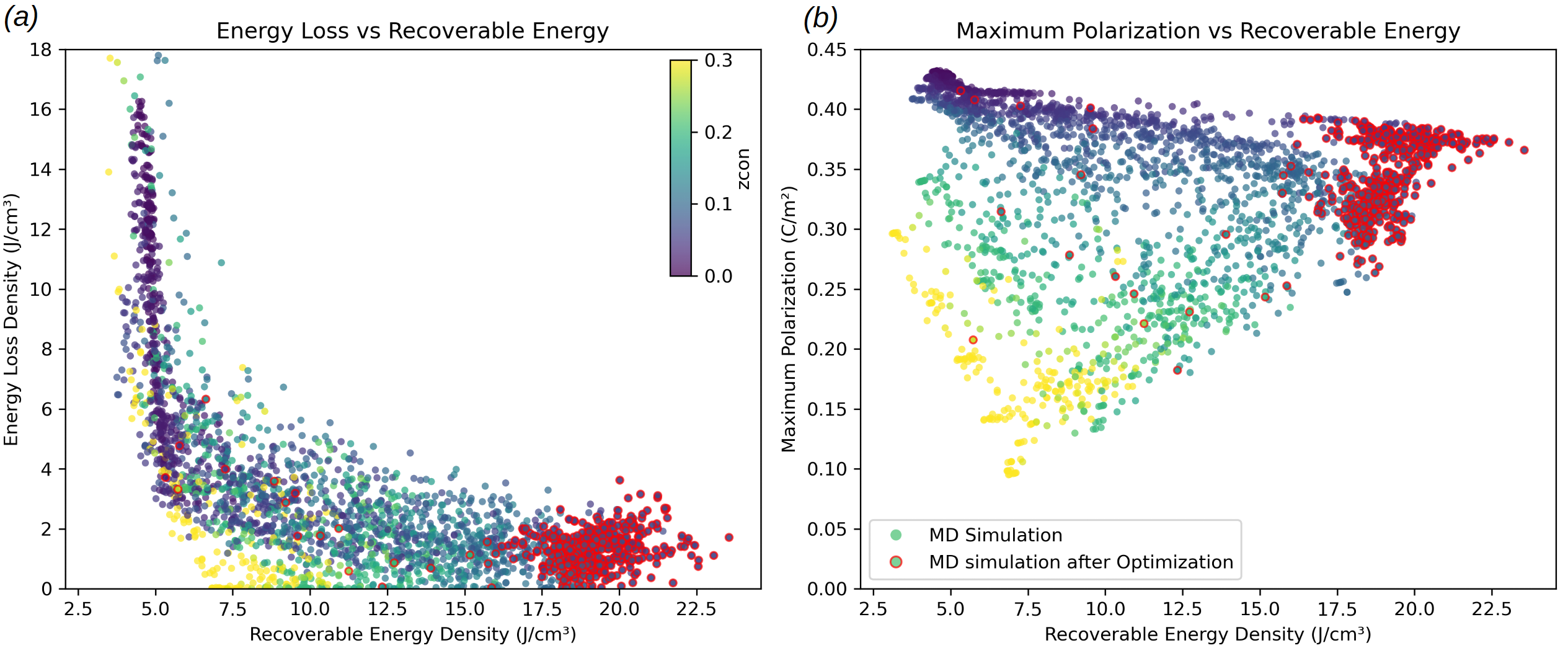}
\caption{Updated EH--MD simulation results (red points) targeted to the ML-identified high-performance region. The additional simulations densely populate the high-$W_\mathrm{rec}$, low-loss regime, confirming that moderate Zr concentrations combined with quasi-continuous layered motifs and a small number of intervals yield improved energy-storage performance.}
\label{fig:optimized MD sim}
\end{figure}

Figure~\ref{fig:optimized MD sim} summarizes the validation results. The figure shows the energy-storage map, i.e., $W_\mathrm{loss}$ versus $W_\mathrm{rec}$ and $P_\mathrm{max}$ versus $W_\mathrm{rec}$, using EH--MD data only and color-coded by Zr concentration (\textit{zcon}). The newly simulated configurations are highlighted in red. The red points populate the high-$W_\mathrm{rec}$ and low-loss region much more densely than the initial sampling, confirming that the surrogate-guided selection drives the search toward improved energy-storage behavior. The $P_\mathrm{max}$--$W_\mathrm{rec}$ projection further shows that most targeted candidates lie in the regime where high $W_\mathrm{rec}$ coincides with high $P_\mathrm{max}$, consistent with the mechanism identified above: high recoverable energy requires both a large polarization amplitude and a narrow hysteresis loop.

\end{multicols}
\vspace{-0.5em}
\begin{multicols}{2}

With this region now more densely sampled and partially validated by EH–MD, the task reduces to selecting a representative configuration that combines high $W_\mathrm{rec}$ with low loss. In our parametrized space, the best-performing candidate corresponds to a layered architecture consisting of BT nanolayers (thickness $\sim$14~nm) periodically interrupted by quasi-continuous Zr-rich layers (thickness $\sim$2~nm). These Zr-rich sheets are BaZrO$_3$-like nanolayers, such that the resulting structure is a simple superlattice along the $z$ direction, which is also the direction of the applied electric field. Although superlattices of ferroelectric and paraelectric materials are not a new concept~\cite{BENYOUSSEF2019279}, this result illustrates the main value of the present workflow: rather than assuming a layered architecture a priori, the approach identifies the relevant distribution features from a broader descriptor space, narrows the search to high-performing regions, and supports targeted atomistic validation.

\subsection{Influence of substitutional disorder on the identified motif–property relationships}\label{sec:random}

\begin{figure}[H]
    \centering
    \includegraphics[width=0.90\linewidth]{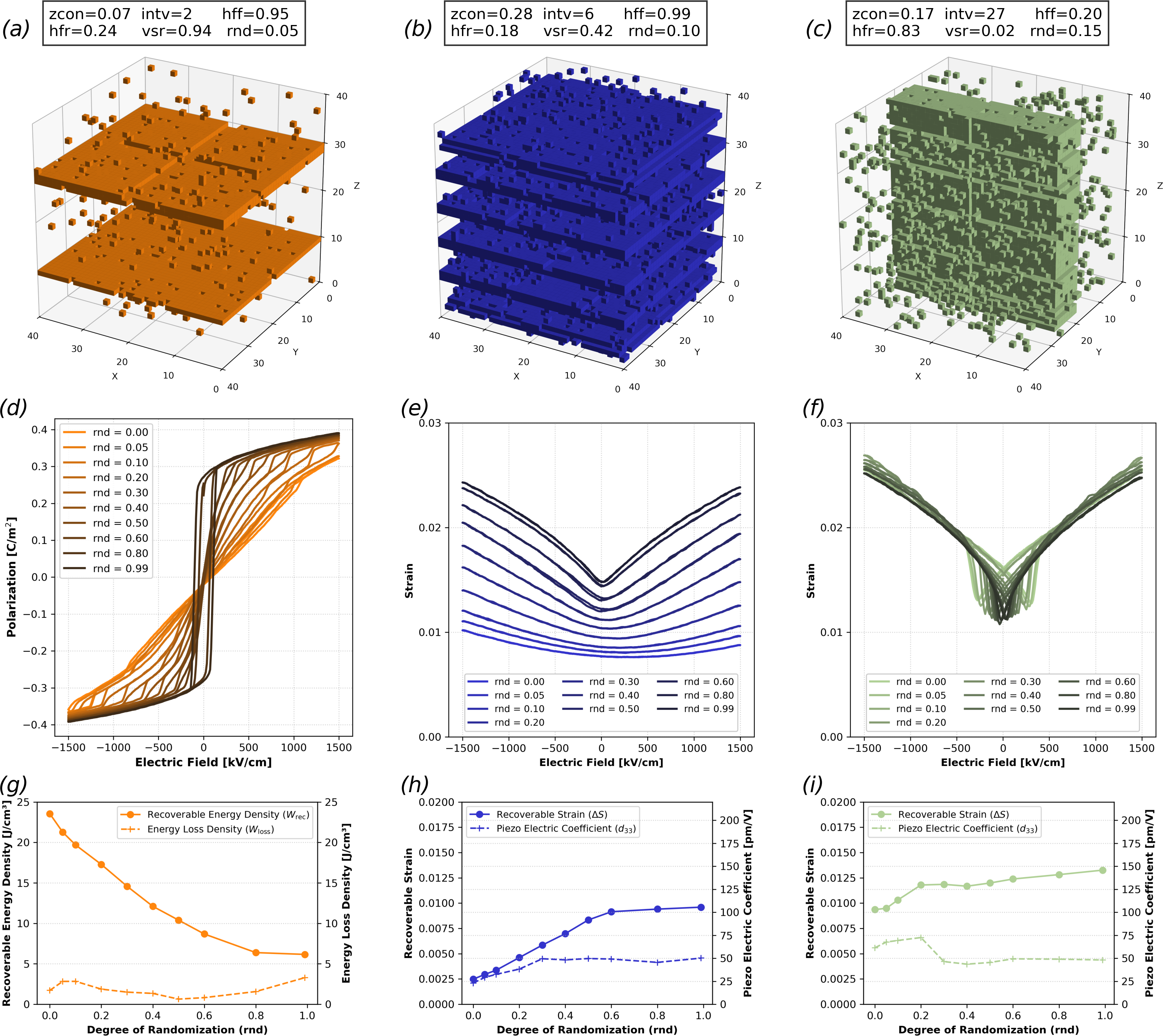}
\caption{\rev{Impact of progressive spatial randomization on the functional response of three representative Zr-distribution motif families corresponding to those identified in Figure~\ref{fig:3dstructures}: a high-energy-storage layered motif (a), a mechanically quiet densely layered motif (c), and a vertical lamella-like motif (e). Panels (a)--(c) show representative three-dimensional realizations of these motifs at blurring ratios $\mathrm{rnd}=0.05$, $0.10$, and $0.15$, respectively, while preserving the overall Zr concentration. Panels (d)--(f) show the evolution of the corresponding response curves over the full investigated range, $\mathrm{rnd}=0.00$--$0.99$: (d) $P$--$E$ loops for the high-energy-storage layered motif and (e),(f) $S$--$E$ loops for the mechanically quiet layered and vertical lamella-like motifs, respectively. Panels (g)--(i) summarize the associated loop-derived quantities as functions of $\mathrm{rnd}$: (g) recoverable energy density $W_{\mathrm{rec}}$ and energy-loss density $W_{\mathrm{loss}}$ for the high-energy-storage motif and (h),(i) recoverable strain $\Delta S$ and high-field effective slope $d_{33}$ for the mechanically quiet layered and vertical lamella-like motifs, respectively. Increasing $\mathrm{rnd}$ progressively weakens the parent structural motif and approaches the disordered solid-solution limit.}}
\label{fig:blurring_robustness}
\end{figure}

\rev{
The primary design space considered above consists of controlled nanoscale Zr-distribution motifs with well-defined interfaces and spatial correlations. Experimentally realized substituted materials, however, may exhibit intermixing, interface roughness, and substitutional disorder. We therefore performed additional direct EH--MD simulations for selected parent structures to examine how progressive randomization of the Zr distribution affects the identified functional regimes.

The degree of disorder is controlled by the blurring parameter \textit{rnd}. Starting from a parent configuration, an increasing fraction of the Zr ions defining the original motif is redistributed among randomly selected B sites while preserving the overall Zr concentration. Accordingly, \textit{rnd}=0 represents the sharply defined parent motif, whereas \textit{rnd}=1 corresponds to a fully randomized solid solution at the same composition. Figure~\ref{fig:blurring_robustness} illustrates the resulting evolution for three representative motif families identified in Figure~\ref{fig:3dstructures}: a high-energy-storage layered motif, a mechanically quiet densely layered motif, and a vertical lamella-like motif characterized by a high $d_{33}$ and moderate $\Delta S$. These examples illustrate three distinct motif-dependent responses to progressive blurring.

The influence of disorder is particularly pronounced for the high-energy-storage layered configuration shown in Figure~\ref{fig:blurring_robustness}a. In the unblurred structure, quasi-continuous Zr-rich layers interrupt the BT matrix perpendicular to the applied field and produce a slim P--E loop with favorable energy-storage characteristics. As \textit{rnd} increases, the sharp compositional modulation is progressively weakened and the corresponding loops evolve toward a more open ferroelectric response, as shown in Figure~\ref{fig:blurring_robustness}d. Both the maximum and remanent polarization increase, while the curvature of the discharge branch becomes less favorable for recoverable energy storage. Consequently, $W_{rec}$ decreases with increasing blurring, as shown in Figure~\ref{fig:blurring_robustness}g. The sharply layered configuration therefore substantially outperforms its randomized counterpart at the same Zr concentration. This confirms that the enhanced energy-storage response does not arise from composition alone, but depends critically on the spatial organization and continuity of the Zr-rich layers.

A similarly clear effect is observed for the mechanically quiet layered configuration shown in Figure~\ref{fig:blurring_robustness}b. The closely spaced Zr-rich layers strongly suppress the field-induced strain response, resulting in an almost flat S--E loop for \textit{rnd}=0. Progressive blurring weakens this dense layered modulation and leads to an increasingly pronounced strain response, as shown in Figure~\ref{fig:blurring_robustness}e. Both the recoverable strain $\Delta S$ and the high-field effective slope $d_{33}$ increase with \textit{rnd} (Figure~\ref{fig:blurring_robustness}h). This behavior is consistent with the loss of the continuous compositional interruptions that constrain the electromechanical response of the BT-rich regions. The mechanically quiet state is therefore promoted by the deliberately structured layering and is gradually lost as the Zr distribution approaches the random limit.

The vertical lamella-like configuration exhibits a different sensitivity to disorder. As shown in Figure~\ref{fig:blurring_robustness}c, this motif combines a moderate recoverable strain with a comparatively high $d_{33}$ and an approximately linear S--E response over a broad field range. Blurring produces only moderate changes in the loop shape and in the derived electromechanical quantities (Figures~\ref{fig:blurring_robustness}f and \ref{fig:blurring_robustness}i), and the response of the fully randomized configuration remains relatively similar to that of the ideal lamella. Unlike the horizontal layers, the vertical Zr-rich feature does not form a continuous compositional barrier perpendicular to the applied field. Connected BT-rich regions therefore remain available to accommodate the field-induced strain in both the lamellar and randomized configurations. The electromechanical response of this motif is consequently less dependent on an atomically sharp interface and more robust against substitutional disorder.

Overall, the blurring analysis demonstrates that substitutional disorder has a substantial but strongly motif-dependent influence on the functional response. Disorder progressively erodes the extreme properties associated with sharply defined horizontal layers: it lowers the recoverable energy density of the optimized energy-storage configuration and increases the strain activity of the mechanically quiet structure. By contrast, the high-$d_{33}$, nearly linear response of the vertical lamella is comparatively insensitive to blurring. These results support the central conclusions of the controlled-motif screening. The spatial distribution of Zr remains a decisive design variable, while disorder primarily modifies the magnitude and sharpness of the targeted response rather than overturning the identified motif--property relationships. This two-stage strategy enables the identification of interpretable structural design rules followed by an independent assessment of their sensitivity to substitutional disorder.}

\end{multicols}
\vspace{-0.5em}
\begin{multicols}{2}

\section{Summary and Conclusions}

We have presented a loop-level workflow for dopant-distribution engineering in substituted ferroelectrics by combining a controlled descriptor model for nanoscale dopant arrangements, effective-Hamiltonian molecular dynamics (EH--MD), and a machine-learning surrogate based on a conditional autoencoder. Using Ba(Zr$_x$Ti$_{1-x}$)O$_3$ (BZT) as a test case, we showed that dopant distribution acts as an additional design variable within the explored descriptor space: at fixed composition, different spatial arrangements of Zr can produce markedly different polarization--electric-field and strain--electric-field hysteresis responses, and therefore different loop-derived figures of merit.

A central outcome of the present work is methodological. Rather than predicting a small set of scalar properties directly, the surrogate model reconstructs complete P--E and S--E loops from compact descriptors of dopant arrangement. This preserves the internal coupling between dielectric and electromechanical response and makes it possible to evaluate multiple derived quantities from a common predicted response representation. Once trained on EH--MD data, the surrogate enables rapid generation of dense response databases at a computational cost far below that of direct simulation, thereby making large-scale screening of dopant-distribution space tractable. Comparison with a direct scalar-regression MLP baseline further shows that the descriptor space is already highly informative, while the cAE remains advantageous because it provides the best overall performance across most targets and, more importantly, predicts complete response curves rather than only a predefined set of scalar outputs.

Within this framework, we constructed property-selective design maps for three representative targets: energy-storage performance, electromechanical response, and switching behavior. Across these cases, the screening results show that distinct motif families occupy different response regimes, demonstrating that nanoscale dopant arrangement can systematically reshape hysteresis behavior beyond trends set by average composition alone. For the energy-storage target, targeted EH--MD follow-up confirmed that moderate Zr concentrations combined with quasi-continuous Zr-rich layering yield slim hysteresis loops with high recoverable energy density, illustrating how surrogate-guided exploration can be translated into validated candidate structures. At the same time, the broader analysis shows that layered motifs are only one part of the accessible design space. Lamellar and nanoplate-like arrangements emerge in other functional regimes. \rev{Furthermore, systematic randomization testing showed that substitutional disorder modifies the functional response in a strongly motif-dependent manner. Progressive blurring reduces the enhanced recoverable energy density of the optimized layered configuration and weakens the suppression of electromechanical activity in the mechanically quiet layer motif, whereas the response of the vertical lamella remains comparatively less sensitive. The disorder analysis therefore supports the identified motif--property relationships while showing that the magnitude of the targeted response depends on the preservation of the underlying spatial correlations.}

The present results should be interpreted within the scope of the adopted model hierarchy. First, the descriptor space is intentionally restricted to controlled motif families and their intermediates, rather than arbitrary disordered substitution patterns. Second, the surrogate is trained and validated primarily for interpolation within the sampled configuration space, so its predictions should be understood in that domain. Third, the EH--MD framework is known to overestimate the electric-field scale quantitatively, meaning that the field-dependent quantities reported here are most meaningful in comparative rather than absolute terms. Accordingly, the main value of the present study lies in identifying trends, rankings, and motif--response relationships under a common simulation protocol, rather than in making direct quantitative predictions for experiment.
\rev{
These limitations also define the most important directions for future work. On the structural side, while our density-preserving randomization model provided a fundamental baseline for spatial disorder, future extensions can incorporate more sophisticated, synthesis-aware blurring mechanisms—such as thermodynamic phase segregation, diffuse Gaussian interdiffusion fronts, or strain-mediated defect clustering—to directly reflect processing-induced interfacial morphologies.} On the surrogate-modeling side, more stringent validation strategies, uncertainty-aware prediction, and active-learning or Bayesian-optimization loops could further strengthen the connection between screening and targeted EH--MD sampling. Finally, the same general framework should be transferable to other substituted ferroic and functional oxide systems in which compact structural descriptors can be linked to field-driven response curves. In this sense, the present study provides not only a case study for BZT, but also a general route for combining physics-based simulation and surrogate modeling to accelerate exploration of complex composition--structure--property landscapes.

\section{Data Availability}
The data supporting the findings of this study are available from the corresponding author upon reasonable request.
\section{Acknowledgments}
The authors gratefully acknowledge the financial support under the scope of the COMET program within
the K2 Center "Integrated Computational Material, Process and Product Engineering (IC-MPPE)"
(Project No 886385). This program is supported by the Austrian Federal Ministries for Economy, Energy
and Tourism (BMWET) and for Innovation, Mobility and Infrastructure (BMIMI), represented by the
Austrian Research Promotion Agency (FFG), and the federal states of Styria, Upper Austria and Tyrol

\printbibliography

\end{multicols}
\newpage
\clearpage
\setcounter{section}{0}
\setcounter{subsection}{0}
\setcounter{subsubsection}{0}
\setcounter{figure}{0}
\renewcommand{\thefigure}{S\arabic{figure}}
\setcounter{page}{1}

\begin{center}
\vspace{4em}
{\LARGE\bfseries Supplementary Material\par}
{\large Loop-level surrogate modeling of dopant-distribution effects in Ba(Zr,Ti)O$_3$\par}
\vspace{1em}
  {\normalsize
  Heiko Röthl\textsuperscript{1},
  Elke Kraker\textsuperscript{1},
  Julien Magnien\textsuperscript{1},
  Manfred Mücke\textsuperscript{1},
  Florian Mayer\textsuperscript{1}\textsuperscript{$\dagger$} \\[0.5em]
  \textsuperscript{1}\,
  Materials Center Leoben Forschung GmbH, Vordernberger Straße 12, 8700 Leoben, Austria \\[0.3em]
  \texttt{\textsuperscript{$\dagger$}florian.mayer@mcl.at}
  }
\vspace{4em}
\end{center}

\section{Details of Supercell Construction}
The parametrized construction of Zr distributions described in Section 2.2 is implemented deterministically once the structural parameters (\textit{zcon}, \textit{intv}, \textit{hff}, \textit{hfr}, \textit{vsr}) are specified.
The following points summarize practical aspects of the generation procedure that are not essential to the conceptual description given in the main text.

\textbf{Layer construction:}
Supercells are filled sequentially layer by layer along the vertical direction.
Active layers are assigned according to the prescribed number of intervals (\textit{intv}), with each interval representing either a single layer or a continuous stack of layers containing Zr substitution.

\textbf{Distribution of substituted sites:} 
Within each active layer, substituted Ti sites are placed inside a single continuous patch whose area corresponds to the horizontal fill factor (\textit{hff}).
Patch dimensions are computed on the discrete lattice grid. When integer rounding prevents exact realization of the target area, remaining sites are assigned to neighboring lattice positions to preserve continuity and maintain the prescribed overall concentration (\textit{zcon}).

\textbf{Patch geometry:}
The in-plane shape of the patch is controlled through the horizontal form ratio (\textit{hfr}).
Aspect ratios are translated to integer lattice dimensions before site assignment.
Minor deviations arising from discretization are absorbed locally and do not affect global parameter consistency.

\textbf{Interval alignment:}
Successive intervals may be laterally shifted according to the vertical superposition ratio (\textit{vsr}).
The shift is applied as a translation vector within the periodic supercell.
Sites translated beyond the boundary are wrapped according to periodic boundary conditions.

\textbf{Parameter consistency:} The algorithm ensures that

\begin{itemize}
    \item the total number of substituted sites matches \textit{zcon}
    \item the number of active intervals matches \textit{intv}
    \item the fill fraction per layer satisfies \textit{hff}
    \item no configuration exceeds the vertical extent of 40 layers
\end{itemize}

Small lattice discretization deviations are corrected during construction to maintain global consistency.
\rev{
To examine the influence of substitutional disorder, an additional blurring parameter \textit{rnd} was introduced. For a parent configuration containing $N_{Zr}$ substituted B sites, a fraction \textit{rnd} of the Zr ions is removed from its original motif positions and redistributed among randomly selected B sites that are occupied by Ti after the removal step. The number of redistributed ions is therefore $N_{blur}$=$round(rndN_{Zr})$. Site selection is performed without replacement, such that every B site remains occupied by either Ti or Zr and the total Zr concentration is conserved exactly. Accordingly, \textit{rnd}=0 retains the original sharp motif, whereas \textit{rnd}=1 produces a fully randomized Zr distribution at the same overall composition. Intermediate values generate partially disordered structures that retain a progressively decreasing fraction of the parent motif. The blurred configurations were evaluated by direct EH--MD simulations and were not included in the training of the five-parameter surrogate model.

Selected parent configurations were evaluated at several blurring ratios using direct EH--MD simulations. These calculations are discussed in Section~\ref{sec:random} of the main text to assess the sensitivity of representative motif--property relationships to substitutional disorder.

Including \textit{rnd} as an additional surrogate input would substantially enlarge the sampling space and require a correspondingly larger number of EH--MD loop sets for adequate training coverage. Because the primary objective was to identify interpretable motif--property relationships and promising ordered structures, \textit{rnd} was instead reserved for the independent disorder-robustness analysis.}

\section{Comparison with a direct scalar-regression baseline}

To benchmark the loop-level conditional autoencoder (cAE) against a simpler alternative, we trained a standard multilayer perceptron (MLP) to predict the principal loop-derived scalar quantities simultaneously from the five structural descriptors $(zcon, intv, hff, hfr, vsr)$. The comparison is summarized in Table~\ref{tab:cae_mlp_comparison} for the validation, test, and full datasets using mean squared error (MSE), root mean squared error (RMSE), and coefficient of determination ($R^2$).

The MLP performs well across all targets, indicating that the adopted descriptor set already captures much of the relevant structure--property mapping within the sampled design space. At the same time, the cAE yields the best overall performance for most quantities, with particularly clear improvements for remanent polarization and the energy-density-related metrics. The MLP remains competitive, and for some localized quantities such as the coercive-field-related target and the high-field effective slope $d_{33}$ it can match or even slightly exceed the cAE on individual splits. This behavior is consistent with the fact that the descriptor space is relatively low-dimensional and informative, so that even direct scalar regression captures a substantial part of the underlying trends.

The main advantage of the cAE is therefore not only predictive accuracy, but also representational flexibility. Unlike the MLP baseline, which must be trained for a predefined set of scalar outputs, the cAE predicts complete P--E and S--E loops. This provides access to the full field-driven response, preserves the coupling between dielectric and electromechanical behavior, and allows additional loop-derived quantities to be evaluated without retraining the model. The baseline comparison thus shows that direct scalar regression is already a strong reference model in the present descriptor space, while the loop-level surrogate remains the more general and physically informative framework.

\begin{table*}[t]
\centering
\caption{Comparison of the loop-level conditional autoencoder (cAE) and a direct scalar-regression multilayer perceptron (MLP) baseline for the principal loop-derived target quantities. Reported metrics are mean squared error (MSE), root mean squared error (RMSE), and coefficient of determination ($R^2$) for the validation, test, and full datasets.}

\vspace{4em}

\label{tab:cae_mlp_comparison}
\small
\setlength{\tabcolsep}{5pt}
\renewcommand{\arraystretch}{1.15}
\begin{tabular}{llrrrrrr}
\toprule
Target & Split & \multicolumn{2}{c}{MSE} & \multicolumn{2}{c}{RMSE} & \multicolumn{2}{c}{$R^2$} \\
\cmidrule(lr){3-4}\cmidrule(lr){5-6}\cmidrule(lr){7-8}
 &  & cAE & MLP & cAE & MLP & cAE & MLP \\
\midrule

\multirow{3}{*}{$P_{\max}$}
& Validation & 2.373e-05 & 4.501e-05 & 0.0049 & 0.0067 & 0.9954 & 0.9930 \\
& Test       & 4.126e-05 & 4.520e-05 & 0.0064 & 0.0067 & 0.9939 & 0.9932 \\
& All        & 2.692e-05 & 3.690e-05 & 0.0052 & 0.0061 & 0.9954 & 0.9936 \\
\midrule

\multirow{3}{*}{$P_{\mathrm{rem}}$}
& Validation & 1.045e-03 & 1.535e-03 & 0.0323 & 0.0392 & 0.9272 & 0.9150 \\
& Test       & 5.784e-04 & 1.351e-03 & 0.0241 & 0.0368 & 0.9638 & 0.9028 \\
& All        & 5.254e-04 & 1.015e-03 & 0.0229 & 0.0319 & 0.9674 & 0.9371 \\
\midrule

\multirow{3}{*}{$E_c$}
& Validation & 7.299e-06 & 7.389e-06 & 0.0027 & 0.0027 & 0.9442 & 0.9582 \\
& Test       & 1.309e-05 & 7.107e-06 & 0.0036 & 0.0027 & 0.9216 & 0.9307 \\
& All        & 7.220e-06 & 5.929e-06 & 0.0027 & 0.0024 & 0.9504 & 0.9592 \\
\midrule

\multirow{3}{*}{$W_{\mathrm{tot}}$}
& Validation & 1.0649 & 1.2658 & 1.0320 & 1.1251 & 0.9421 & 0.9288 \\
& Test       & 1.1244 & 1.3245 & 1.0604 & 1.1509 & 0.9341 & 0.9116 \\
& All        & 0.8564 & 0.9832 & 0.9254 & 0.9915 & 0.9478 & 0.9401 \\
\midrule

\multirow{3}{*}{$W_{\mathrm{rec}}$}
& Validation & 0.6872 & 0.7356 & 0.8290 & 0.8577 & 0.9763 & 0.9753 \\
& Test       & 0.6334 & 1.1941 & 0.7959 & 1.0928 & 0.9759 & 0.9516 \\
& All        & 0.5693 & 0.7461 & 0.7545 & 0.8638 & 0.9796 & 0.9733 \\
\midrule

\multirow{3}{*}{$W_{\mathrm{loss}}$}
& Validation & 0.6941 & 0.8720 & 0.8331 & 0.9338 & 0.9447 & 0.9475 \\
& Test       & 1.0326 & 0.9470 & 1.0162 & 0.9731 & 0.9376 & 0.8928 \\
& All        & 0.6243 & 0.6537 & 0.7901 & 0.8085 & 0.9544 & 0.9522 \\
\midrule

\multirow{3}{*}{$S_{\max}$}
& Validation & 2.812e-07 & 5.642e-07 & 5.303e-04 & 7.511e-04 & 0.9895 & 0.9828 \\
& Test       & 4.719e-07 & 6.294e-07 & 6.870e-04 & 7.934e-04 & 0.9872 & 0.9812 \\
& All        & 3.049e-07 & 4.767e-07 & 5.522e-04 & 6.905e-04 & 0.9902 & 0.9847 \\
\midrule

\multirow{3}{*}{$d_{33}$}
& Validation & 252.0083 & 206.9774 & 15.8748 & 14.3867 & 0.8288 & 0.8106 \\
& Test       & 308.1278 & 240.1141 & 17.5536 & 15.4956 & 0.7209 & 0.7855 \\
& All        & 135.0556 & 176.5903 & 11.6213 & 13.2887 & 0.8750 & 0.8366 \\
\bottomrule
\end{tabular}
\end{table*}

\section{Model Development and Training Details}

The overarching objective of the surrogate-model development was to establish a reliable inference pathway that maps structural descriptors of Zr distributions directly to electromechanical response sequences. While sequence reconstruction forms the technical foundation of the approach, the ultimate purpose of the modeling effort is predictive: enabling rapid estimation of configuration-dependent polarization– and strain–field behavior without performing additional EH–MD simulations.

Model development therefore proceeded in staged fashion. Initial efforts focused on verifying that the response sequences themselves could be compactly represented and reconstructed by neural autoencoders. Only after this feasibility had been demonstrated was the architecture extended toward conditional inference capable of linking structural inputs to response outputs. The subsections below describe this progression.

\subsection{Development Strategy and Initial Feasibility Testing}

Development of the surrogate model followed a gradual bottom-up strategy aimed at first establishing the suitability of sequence autoencoding for representing electromechanical response loops before introducing structural conditioning. The initial objective was therefore not prediction from structural descriptors, but verification that polarization–electric field (P--E) and strain–electric field (S--E) sequences could be reconstructed with sufficient fidelity using a compact neural representation.

As a first step, a baseline autoencoder was constructed using only polarization–field loops as input and target. This simplified setting allowed systematic evaluation of architectural choices such as convolutional depth, dense-layer size, and compression level without the additional complexity introduced by conditioning. Numerous variants were tested to identify configurations capable of capturing switching behavior, loop curvature, and hysteretic asymmetry while remaining numerically stable during training.

These feasibility tests revealed important characteristics of the sequence learning task. 
Strongly ferroelectric loops with steep switching gradients consistently represent one challenging regime: while overall loop shapes were reproduced reliably, slight overshoot following rapid transitions was frequently observed. This behavior persisted across architectural variants and therefore reflects an intrinsic difficulty of representing sharp nonlinear transitions using discretized convolutional sequence models rather than a consequence of conditioning or latent-space design.

A second class of challenging responses consists of complex relaxor-like loops characterized by moderate loss, multiple curvature changes, and localized kinks along the field axis. Unlike strongly switching loops, these responses do not exhibit dominant global structure; instead, they combine smooth and sharply varying segments distributed across the sequence. Such patterns are inherently less compressible within a low-dimensional latent representation and therefore more difficult to reconstruct accurately.

As the dataset expanded during later stages of model development, reconstruction of these responses appeared to become more stable and less prone to localized deviations. This improvement was observed qualitatively during iterative experimentation; however, no controlled ablation study isolating dataset size effects was performed. The observation is therefore reported here as an empirical trend rather than a quantified result.

Together with the behavior observed for steep switching loops, these findings provided practical guidance for subsequent model development and established realistic expectations regarding reconstruction behavior prior to introducing conditional inputs.

Once reconstruction capability for a single response sequence had been established, the autoencoder was extended to jointly encode polarization and strain loops. Because both responses originate from the same atomistic simulations and are physically coupled, simultaneous modeling allowed the network to exploit shared electromechanical structure. This stage confirmed that multi-sequence compression could be achieved without degradation of reconstruction stability and provided the basis for the subsequent introduction of structural conditioning described in the following sections.

\newpage
\subsection{Conditional Inference Architecture}

The central objective of the surrogate model is the construction of a reliable inference pathway that predicts electromechanical response sequences from structural descriptors. Rather than relying on a single transformation from structural descriptors to sequence, the adopted architecture establishes two parallel information pathways that pursue the same reconstruction objective.

One pathway encodes the response sequences themselves, while a second processes the structural descriptors defining the Zr distribution. Both encoders map their inputs into latent representations of identical dimensionality. This design enables direct comparison of the representations and establishes a shared embedding space in which structural information and response characteristics can be aligned.

Consistency between the two representations is encouraged through an additional network output, here referred to as the zero head. This branch evaluates the difference between the latent embeddings and is trained against a fixed zero target, thereby penalizing discrepancies between sequence-derived and structure-derived representations. The resulting constraint promotes convergence toward mutually compatible embeddings without forcing either pathway to dominate the representation.

Once aligned, latent features are propagated through a shared decoder that reconstructs polarization and strain sequences simultaneously. This ensures that both inference pathways ultimately contribute to sequence prediction while preserving the physical coupling between the two response types.

Architectural variants explored during development primarily concerned decoder structure and strategies for incorporating conditional information while maintaining the dual-path alignment mechanism described above. Both shared-weight decoders and separate decoder branches were implemented and evaluated. No systematic differences in reconstruction stability or predictive behavior were observed between these alternatives once conditional pathways were sufficiently established.

The placement of conditional inputs had a more noticeable influence. Structural descriptors were injected into the sequence encoder and reintroduced during decoding. Conditioning the encoder encourages structural information to become embedded directly within the latent representation, facilitating alignment between sequence-derived and structure-derived embeddings and reducing corrective adjustments required from the zero-head constraint.

Reinjection of conditional information during decoding serves a complementary purpose by preventing the latent representation from becoming overloaded with structural information. This allows latent capacity to remain available for representing sequence-specific features while still preserving structural guidance during reconstruction.

Conditional features were reinjected during decoding either directly at the latent level or after intermediate dense transformations. Both approaches yielded comparable behavior, and the latter was adopted for the final architecture for implementation consistency.

\subsection{Loss Function Formulation}

Training of the conditional autoencoder employed two loss measures serving distinct purposes. 
Mean squared error (MSE) was used consistently as the underlying error metric across all components. 
Model parameters were updated using a composite training loss during backpropagation, whereas a separate validation loss assessed generalization of the structural-to-sequence inference pathway.

The training objective combines three contributions.  
First, reconstruction of the sequence autoencoder branch ensures stable learning of compact representations of the hysteresis responses.  
Second, the conditional inference pathway is supervised by enforcing agreement between sequences predicted from structural descriptors and the corresponding simulation data.  
Finally, consistency between latent representations obtained from the sequence and conditional encoders is encouraged through a penalty acting on their difference.  
Relative influence of the conditional and latent-consistency terms is controlled through weighting factors.

Validation performance was assessed independently of the composite objective.  
Because the intended use of the surrogate model is prediction of response sequences from structural descriptors, the validation metric evaluates only the conditional sequence prediction accuracy.  
This prevents improvements in auxiliary reconstruction tasks from biasing model selection and ensures that generalization reflects inference capability.

The resulting loss definitions are

\begin{align}
L_{\mathrm{train}} &= 
L_{\mathrm{seq\to seq}}
+ \lambda_{\mathrm{cond}} L_{\mathrm{cond\to seq}}
+ \lambda_{\mathrm{latent}} L_{\mathrm{latent}} \\[6pt]
L_{\mathrm{seq\to seq}} &= \mathrm{MSE}(S_{\mathrm{AE}}, S) \\[4pt]
L_{\mathrm{cond\to seq}} &= \mathrm{MSE}(S_{\mathrm{cond}}, S) \\[4pt]
L_{\mathrm{latent}} &= \mathrm{MSE}(z_{\mathrm{seq}} - z_{\mathrm{cond}}, 0) \\[6pt]
L_{\mathrm{val}} &= \mathrm{MSE}(S_{\mathrm{cond}}, S)
\end{align}

where $S$ denotes the reference sequences obtained from EH–MD simulations,  
$S_{\mathrm{AE}}$ and $S_{\mathrm{cond}}$ represent reconstructions from the autoencoder and conditional pathways respectively, and $z_{\mathrm{seq}}$ and $z_{\mathrm{cond}}$ are their corresponding latent embeddings.  $\lambda_{\mathrm{cond}}$ and $\lambda_{\mathrm{latent}}$ denote weighting factors.

The formulation above represents the baseline objective used throughout model development. 
While this composite loss already yields stable convergence, its effectiveness depends strongly on the relative weighting of the individual contributions. Arbitrary choices of the weighting factors $\lambda_{\mathrm{cond}}$ and $\lambda_{\mathrm{latent}}$ do not prevent convergence, but they lead to suboptimal predictive performance. In particular, insufficient weighting of the conditional pathway reduces inference accuracy, whereas excessive weighting can degrade latent representational structure. The full benefit of the architecture therefore becomes evident only after systematic optimization of these coefficients, as discussed in Section~\ref{sec:cmp_lossvariants}.

\subsection{Loss Function Variants and Physics-Guided Extensions}

In addition to the baseline objective described in the main text, several extended loss formulations were explored during model development. These variants were motivated by systematic discrepancies observed in derived functional descriptors and were designed to bias training toward improved physical consistency of reconstructed response sequences.

\subsubsection{Descriptor-Augmented Loss}

As illustrated by the prediction–reference cross-plots in Figure~\ref{fig:pred-actula-xplot} (main text), remanent polarization and the piezoelectric coefficient show larger scatter than other descriptors. This behavior motivated exploration of descriptor-augmented loss variants during model development. Both quantities are extracted from localized regions of the hysteresis sequences — the field-axis intersection and the high-field slope, respectively —and are therefore sensitive to small reconstruction errors that are only weakly penalized by global sequence MSE.

To address this, an augmented objective was introduced by incorporating explicit penalties on descriptor mismatch. 
Let $L_{\mathrm{base}}$ denote the baseline training objective defined above. 
The descriptor-augmented loss is

\begin{align}
L_{\mathrm{desc}} =
L_{\mathrm{base}}
+ \lambda_{P_r}\, \mathrm{MSE}(P_r^{\mathrm{pred}}, P_r^{\mathrm{ref}})
+ \lambda_{d}\, \mathrm{MSE}(d^{\mathrm{pred}}, d^{\mathrm{ref}})
\end{align}

where

- $P_r$ denotes remanent polarization extracted from the polarization channel  
- $d$ denotes the piezoelectric coefficient  
- $\lambda_{P_r}$ and $\lambda_d$ control the influence of the additional penalties  

These terms bias optimization toward improved agreement in localized sequence-derived descriptors while retaining the dominant sequence-reconstruction objective.

\subsubsection{Energy-Regularized Loss}

In addition to the baseline objective and descriptor-augmented formulation, a further loss variant was explored that biases training toward preserving global energetic characteristics of the hysteresis response. The motivation for this extension was to encourage consistency in loop magnitude and discharge behavior, which directly influence recoverable-energy–related descriptors used in the main analysis.

Rather than computing explicit numerical integrals of the loop area during training, global energy proxies were derived directly from the polarization sequences predicted by the conditional branch. These proxies were constructed from averaged polarization values evaluated over portions of the electric-field cycle corresponding to charging and discharging behavior. They therefore capture relative differences in loop geometry and amplitude while remaining computationally inexpensive to evaluate during optimization.

The resulting terms do not impose absolute thermodynamic energy constraints — no physical scaling to real energy density is applied — but act as structural regularizers that guide the model toward improved preservation of large-scale loop characteristics relevant for energy-storage analysis.

The extended objective augments the baseline training loss with weighted penalties on discrepancies between predicted and reference proxy quantities. This formulation was evaluated alongside other loss variants during model development and used selectively when improved energetic consistency was desired.

\paragraph{Loss formulation}

Let $S$ denote the reference response sequences and $P$ the polarization channel of $S$. 
Predicted sequences obtained from structural descriptors are denoted by $S_{\mathrm{cond}}$ and $P_{\mathrm{cond}}$.

Proxy measures representing global charging and discharging behavior are defined as

\begin{align}
W_{\mathrm{ch}}^{\mathrm{ref}} &= \Phi(P), \\
W_{\mathrm{ch}}^{\mathrm{pred}} &= \Phi(P_{\mathrm{cond}}), \\
W_{\mathrm{dis}}^{\mathrm{ref}} &= \Psi(P), \\
W_{\mathrm{dis}}^{\mathrm{pred}} &= \Psi(P_{\mathrm{cond}})
\end{align}

where $\Phi(\cdot)$ and $\Psi(\cdot)$ denote averaging operators applied to sequence segments associated with charging and discharging portions of the cycle.

The corresponding penalties are

\begin{align}
L_{\mathrm{ch}} &= 
\mathrm{MSE}\!\left(
W_{\mathrm{ch}}^{\mathrm{pred}}, 
W_{\mathrm{ch}}^{\mathrm{ref}}
\right) \\[4pt]
L_{\mathrm{dis}} &= 
\mathrm{MSE}\!\left(
W_{\mathrm{dis}}^{\mathrm{pred}}, 
W_{\mathrm{dis}}^{\mathrm{ref}}
\right)
\end{align}

and the resulting energy-regularized objective becomes

\begin{align}
L_{\mathrm{energy}} =
L_{\mathrm{base}}
+ \lambda_{\mathrm{ch}} L_{\mathrm{ch}}
+ \lambda_{\mathrm{dis}} L_{\mathrm{dis}}
\end{align}

where $\lambda_{\mathrm{ch}}$ and $\lambda_{\mathrm{dis}}$ control the influence 
of the energetic regularization relative to the baseline composite loss.

The augmented objectives defined above were implemented as optional extensions of the baseline loss and used selectively during model development. Their formulation is presented here for completeness; their role in training and hyperparameter tuning is discussed in the following subsection.

\subsection{Training Procedure}

Model training was carried out using the Adam optimizer~\cite{Kingma2015}. Early stopping with adjustable patience was employed to terminate training once validation performance ceased to improve, thereby preventing unnecessary optimization cycles while maintaining generalization stability.

The dataset was randomly partitioned into training, validation, and test subsets using fractions of 0.8, 0.1, and 0.1, respectively. The training subset was used for parameter optimization, the validation subset for early-stopping and hyperparameter selection, and the held-out test subset for final performance evaluation. Random number generators governing dataset partitioning and weight initialization were seeded explicitly to ensure deterministic reproducibility across training runs.

Depending on the weighting parameters of the composite loss function, convergence was typically reached between 80 and 150 epochs. Across all tested configurations the optimization process exhibited stable behavior. Trial runs without early stopping did not show systematic increases in validation loss even beyond 200 epochs, indicating well-conditioned training dynamics and the absence of pronounced overfitting tendencies. These observations suggest that the architecture and loss formulation provide a stable optimization landscape for the considered dataset size and sequence complexity.

In addition to standard parameter optimization, model hyperparameters were systematically tuned. 
Because multi-head architectures exhibit strong sensitivity to architectural and loss-weight choices, explicit optimization was required to obtain stable predictive performance.

A key structural parameter investigated during optimization was the dimensionality of the latent representation. 
This parameter was explored across a range of candidate values and was found to converge consistently toward an optimal size of seven latent dimensions. 
Notably, this result was largely independent of the specific loss formulation employed, suggesting that it reflects intrinsic compressibility of the hysteresis response representation rather than loss-dependent effects.

Hyperparameter optimization proceeded in two stages. 
First, a coarse grid search was conducted to identify promising regions of parameter space, including loss weighting factors, learning rates, and architectural parameters. 
Subsequently, Bayesian optimization was applied within the restricted search region to refine parameter selection. 
Each stage involved approximately 300 training runs, allowing efficient exploration while maintaining manageable computational cost.

The final models selected for analysis represent the best-performing configurations obtained from this optimization procedure. Two competing solutions emerged, both employing the shared-decoder architecture but differing in loss formulation: one using descriptor-augmented supervision and the other employing energy-based regularization. Their training behavior and predictive performance are compared in the following subsection.

\subsubsection{Comparison of Loss Variants} \label{sec:cmp_lossvariants}

\begin{figure}[H]
    \centering
    \includegraphics[width=0.75\linewidth]{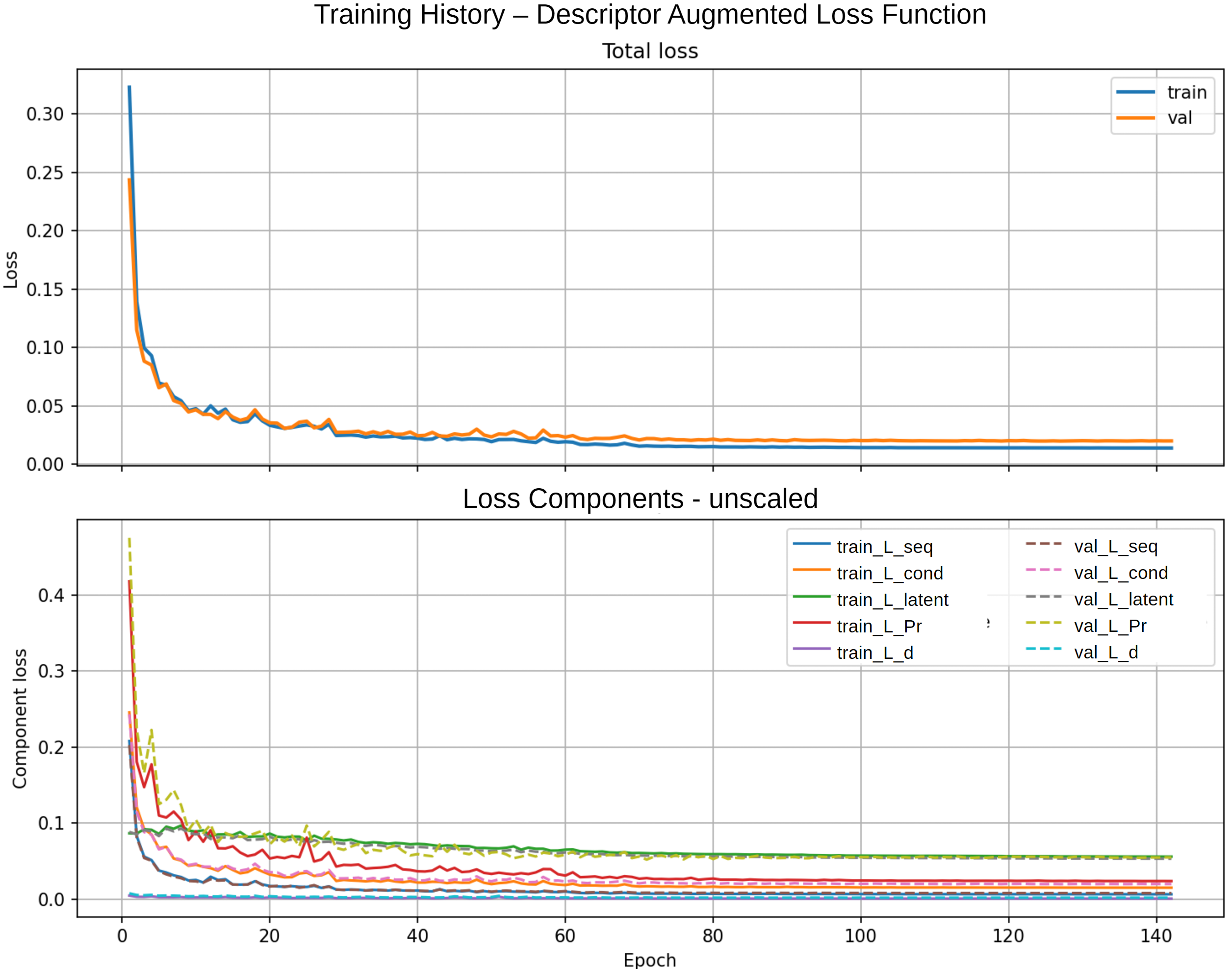}
    \caption{Training history for the conditional autoencoder using the descriptor-augmented loss formulation. Top: total training and validation loss. Bottom: unscaled contributions of individual loss components — sequence reconstruction ($L_{\mathrm{seq}}$), conditional inference ($L_{\mathrm{cond}}$), latent alignment ($L_{\mathrm{latent}}$), and descriptor penalties for remanent polarization ($L_{P_r}$) and piezoelectric coefficient ($L_{d}$). Solid and dashed lines indicate training and validation values, respectively.}
    \label{fig:loss_augmented}
\end{figure}

\begin{figure}[H]
    \centering
    \includegraphics[width=0.75\linewidth]{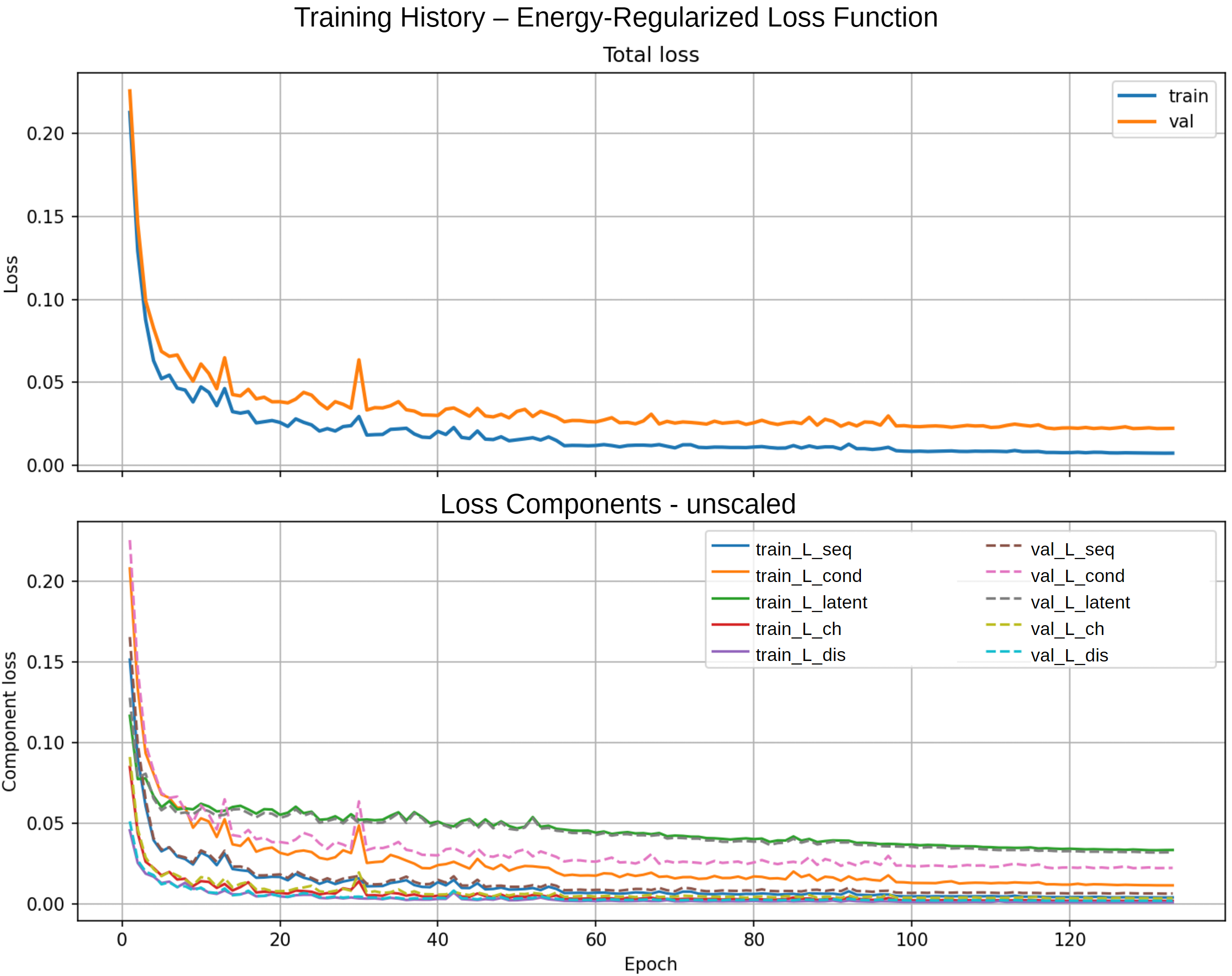}
    \caption{Training history obtained using the energy-regularized loss formulation. The upper panel shows total training and validation losses. The lower panel displays the individual unscaled loss components, including sequence reconstruction, conditional inference, latent alignment, and the proxy-based charging and discharging energy terms.}
    \label{fig:loss_energy}
\end{figure}

Representative training histories obtained using the descriptor-augmented and energy-regularized loss formulations are shown in Figures \ref{fig:loss_augmented}  and \ref{fig:loss_energy}, respectively. In both cases, stable convergence is observed for the total loss as well as for individual loss components, indicating well-conditioned optimization behavior.

Despite their different physical motivations, both objectives lead to very similar predictive performance. The final validation mean-squared-error (MSE) reaches 0.0199 for the descriptor-augmented formulation and 0.0221 for the energy-regularized variant. Visual inspection of reconstructed hysteresis sequences likewise shows comparable reproduction quality for both models.

Small variations in the final validation error were observed to depend on the random initialization seed, and in some instances the relative ordering of the two formulations was reversed. This indicates that the performance difference between the objectives is of the same magnitude as the variability introduced by stochastic training.

A systematic statistical comparison based on multiple repeated training runs was not conducted, as the purpose of these experiments was not benchmarking loss functions but evaluating robustness of the surrogate modeling framework. The observed similarity nevertheless demonstrates that the predictive capability of the model is not strongly sensitive to the specific physics-guided loss variant employed.

\subsection{Implementation and Reproducibility Details}

All machine-learning models were implemented using the PyTorch framework. 
Training data and simulation metadata were organized using an SQLite database backend to enable structured retrieval and experiment tracking. 

Model architectures, loss definitions, and training procedures were implemented in modular components, allowing independent modification of network structure and optimization objectives. 
Runtime configuration — including hyperparameters, dataset selection, and logging options — was controlled through external configuration files to ensure transparent experiment reproducibility.

Unless otherwise noted, training runs were executed on GPU-accelerated hardware. 
Random seeds governing initialization, data shuffling, and partitioning were fixed to ensure deterministic reproducibility of reported results.

\section{Complete descriptor correlation maps for property-selected subsets}
The following figures provide the complete set of pairwise correlations between the five distribution descriptors (\textit{zcon}, \textit{intv}, \textit{hff}, \textit{hfr}, \textit{vsr}) for the 50,000 cAE-predicted configurations. For consistency with the main text, the same percentile-based filters used in Figure~\ref{fig:correlation} are applied to highlight property-selected subsets for energy-storage, electromechanical response, and switching behavior. The purpose of these plots is to document the full correlation structure underlying the selected subsets and to allow identification of descriptor combinations associated with each screened regime. Unless stated otherwise, all percentiles are computed over the full predicted dataset.

\begin{figure}[H]
    \centering
    \includegraphics[width=0.90\linewidth]{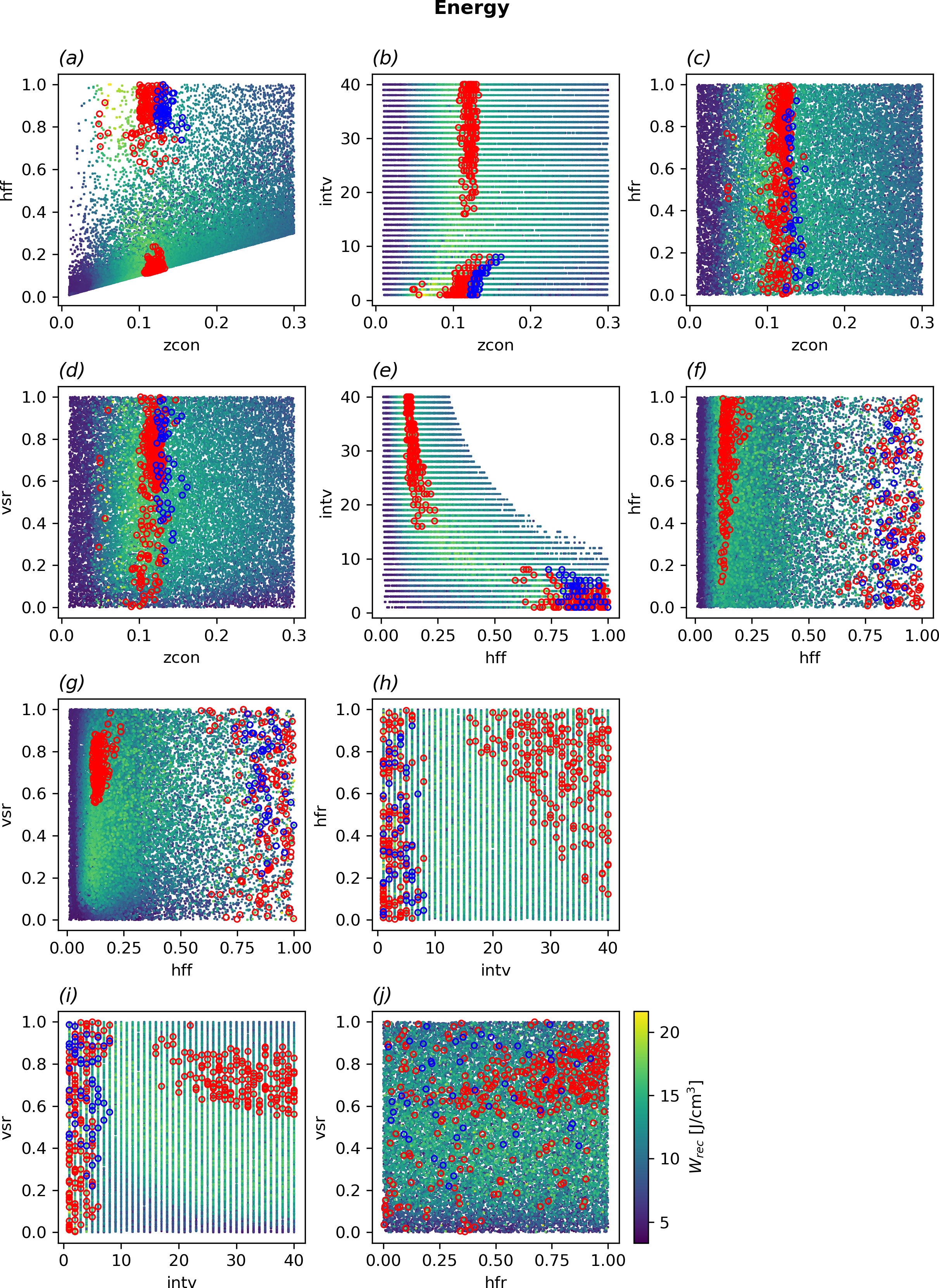}
    \caption{Complete correlation matrix between the five distribution descriptors (\textit{zcon}, \textit{intv}, \textit{hff}, \textit{hfr}, \textit{vsr}) for the cAE-predicted loop database (50,000 configurations), shown for the energy-storage screening. Each point corresponds to one predicted configuration. Red points indicate the high energy-storage subset ($W_\mathrm{rec}\ge$ 90th percentile and $W_\mathrm{loss}\le$ 5th percentile). Blue points indicate the mechanically quiet subset within the red selection \rev{($\Delta S\le$} 30th percentile and $d_{33}\le$ 30th percentile). All percentiles are computed over the full predicted dataset.}
    \label{fig:suppl_energy_correlation}
\end{figure}
\vspace{-0.5em}

\begin{figure}[H]
    \centering
    \includegraphics[width=0.90\linewidth]{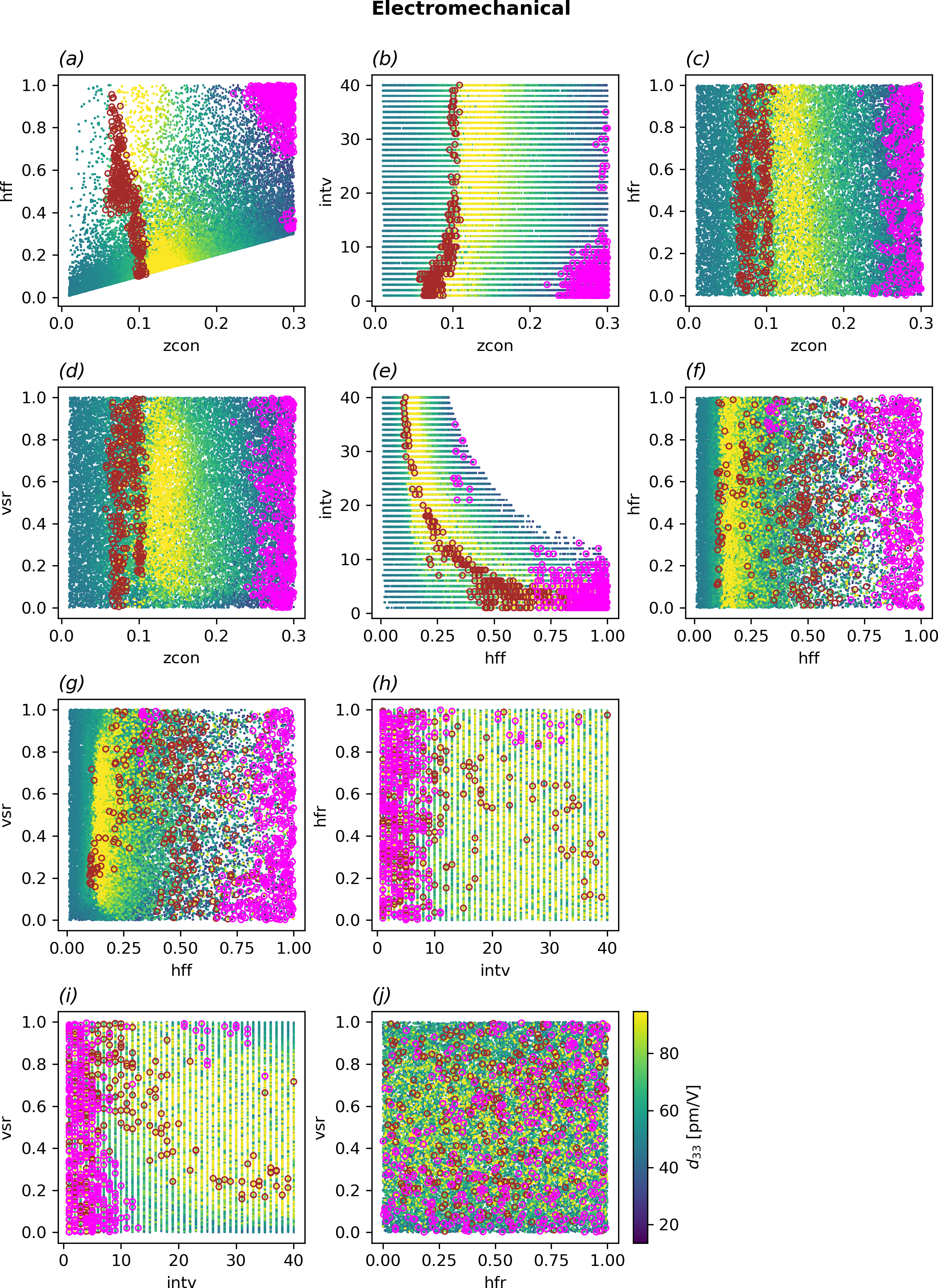}
    \caption{\rev{Complete correlation matrix between the five distribution descriptors (\textit{zcon}, \textit{intv}, \textit{hff}, \textit{hfr}, \textit{vsr}) for the cAE-predicted loop database (50,000 configurations), shown for electromechanical screening. Each point corresponds to one predicted configuration. Dark red points indicate the high-electromechanical subset ($d_{33}\ge$ 85th percentile and $\Delta S\ge$ 85th percentile). Purple points indicate mechanically quiet configurations ($d_{33}\le$ 2nd percentile and $\Delta S\le$~2nd~percentile). All percentiles are computed over the full predicted dataset.}}
    \label{fig:suppl_actuator_correlation}
\end{figure}
\vspace{-0.5em}

\begin{figure}[H]
    \centering
    \includegraphics[width=0.90\linewidth]{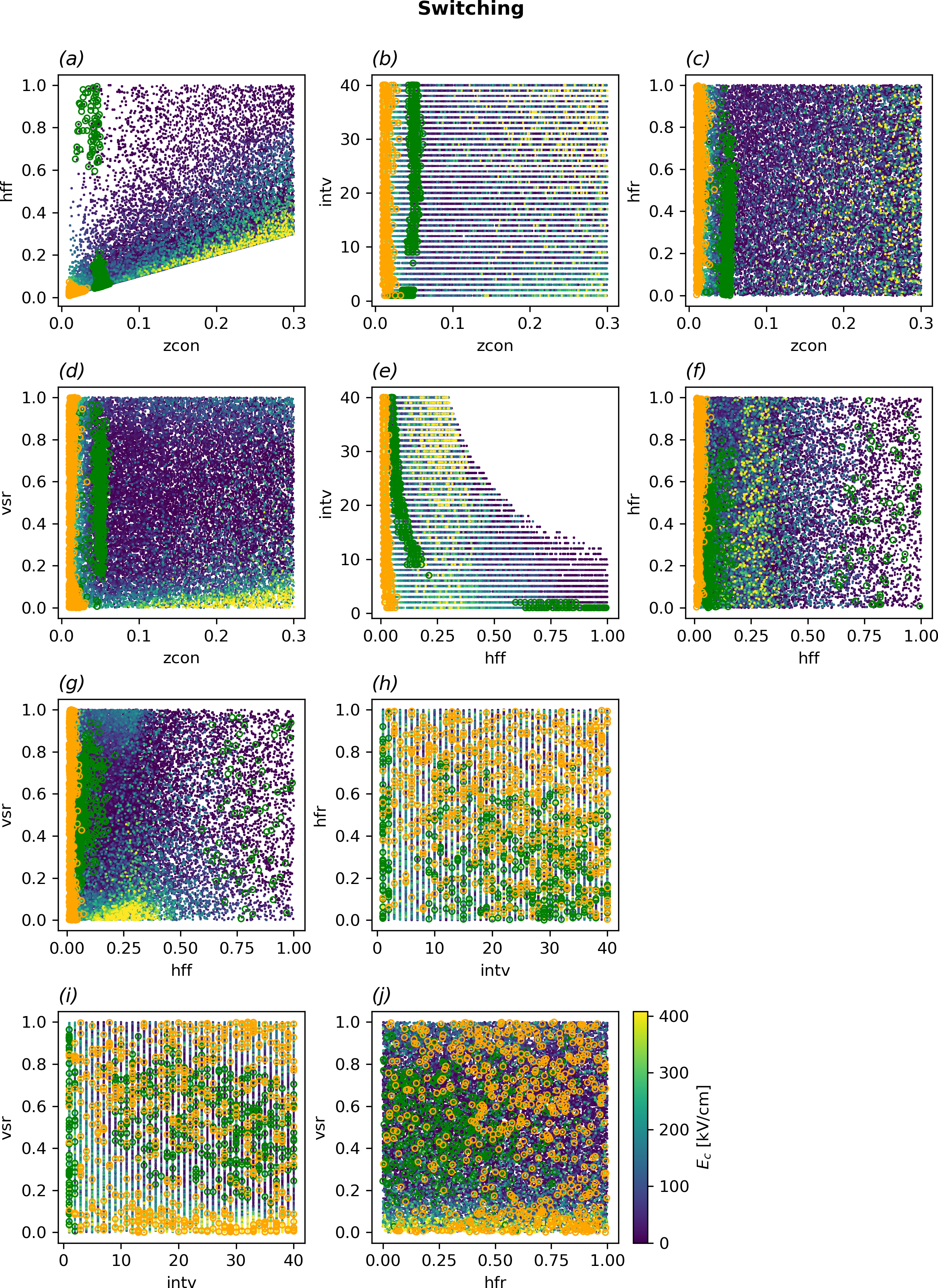}
    \caption{Complete correlation matrix between the five distribution descriptors (\textit{zcon}, \textit{intv}, \textit{hff}, \textit{hfr}, \textit{vsr}) for the cAE-predicted loop database (50,000 configurations), shown for switching-related screening. Each point corresponds to one predicted configuration. Green points indicate the easy-switching subset (low coercive field and low remanent polarization with high maximum polarization: $E_c\le$ 15th percentile, $P_r\le$ 15th percentile, and $P_\mathrm{max}\ge$ 80th percentile). Orange points indicate the hard-switching subset ($E_c\ge$ 95th percentile and $P_r\ge$ 95th percentile). All percentiles are computed over the full predicted dataset.}
    \label{fig:suppl_switching_correlation}
\end{figure}

\vspace{-0.5em}

\section{Additional EH-MD validation}
\begin{figure}[H]
    \centering
    \includegraphics[width=0.95\linewidth]{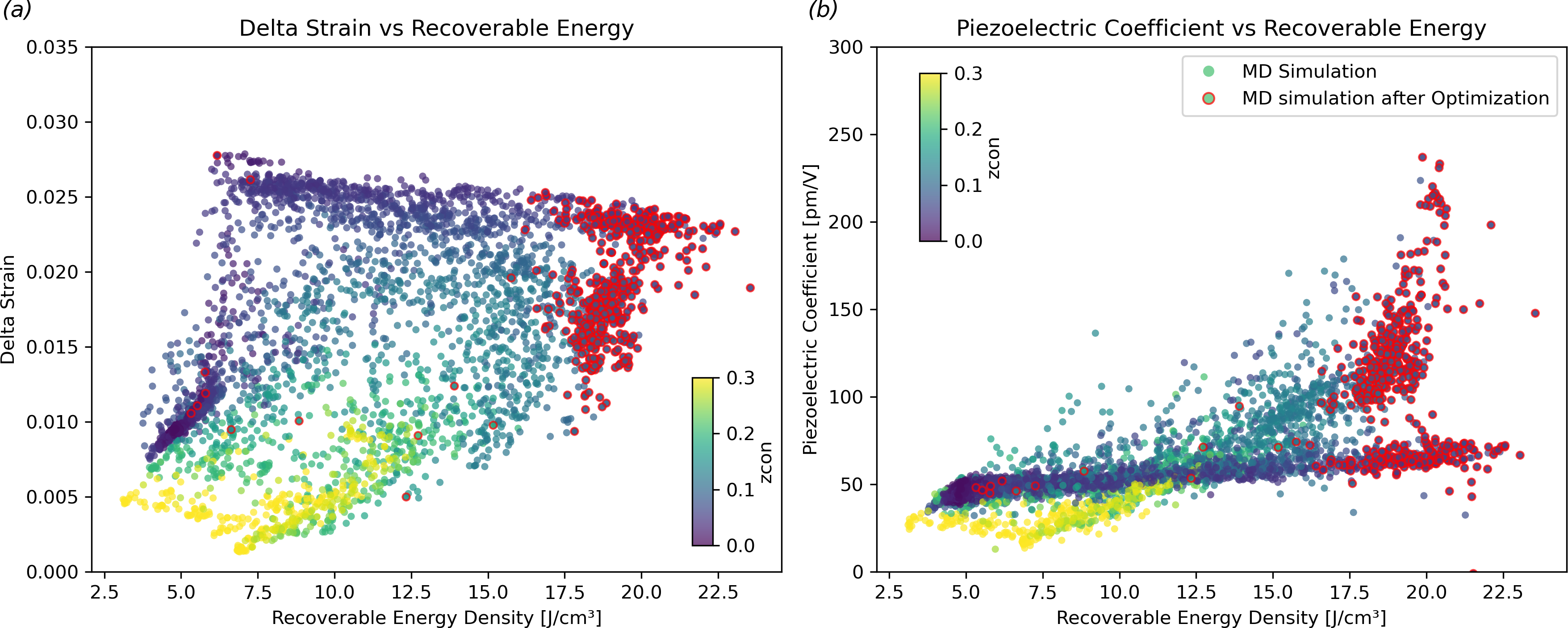}
    \caption{Additional EH--MD validation of the electromechanical screening shown as (a) \rev{Recoverable Strain $(\Delta S)$ and (b) high-field effective slope $(d_{33})$ versus recoverable energy density $(W_{\mathrm{rec}})$. Filled points denote the original EH--MD dataset, color-coded by Zr concentration $(zcon)$, while red open circles indicate additional EH--MD simulations performed for surrogate-selected candidates. Panel (a) shows that the optimized configurations occupy a similar high-$W_{\mathrm{rec}}$ region with broadly comparable $\Delta S$.} Panel (b) reveals a clear branching in $d_{33}$ at comparable $W_{\mathrm{rec}}$, indicating that similar recoverable energy densities and strain amplitudes can arise from distinct electromechanical response modes. Consistent with the motif analysis in the main text, the higher-$d_{33}$ branch is associated mainly with vertical lamella-like motifs and closely related shifted vertical nanoplate variants, whereas the lower-$d_{33}$ branch is dominated by layer-like configurations.}
    \label{fig:fig_S6}
\end{figure}

The additional EH--MD validation shown in Figure~\ref{fig:fig_S6} reveals a clear branching behavior when the high-field effective slope $d_{33}$ is plotted against recoverable energy density $W_{\mathrm{rec}}$, whereas the corresponding \rev{ recoverable strain $\Delta S$ remains broadly comparable} across the optimized configurations. This result is consistent with the electromechanical screening discussed in the main text and provides a more detailed view of the underlying motif dependence. In particular, the higher-$d_{33}$ branch is associated predominantly with vertically oriented lamella-like motifs, as well as slightly shifted vertical nanoplate configurations that can be interpreted as a disrupted variant of the ideal vertical lamella. By contrast, the lower-$d_{33}$ branch is dominated by layer-like configurations. The EH--MD follow-up therefore supports the interpretation that similar high-field strain amplitudes can be achieved through structurally distinct response regimes, which differ primarily in the high-field slope of the S--E loop rather than in the final strain amplitude itself.
\vspace{-0.5em}

\end{document}